\documentclass[twocolumn,amsmath,nofootinbib,amssymb]{revtex4}

\usepackage{epsfig}
\usepackage{graphicx}
\usepackage{dcolumn}
\usepackage{bm}
\usepackage{color}

\newcommand{\lsim}{\mbox{\raisebox{-.9ex}{~$\stackrel{\mbox{$<$}}{\sim}$~}}}
\newcommand{\gsim}{\mbox{\raisebox{-.9ex}{~$\stackrel{\mbox{$>$}}{\sim}$~}}}

\def\thebiblio#1{
\begin{center}\bf \large References
\end{center}
\list
{[\arabic{enumi}]}{\settowidth\labelwidth{#1.}\leftmargin\labelwidth
 \advance\leftmargin\labelsep
 \usecounter{enumi}}
 \def\newblock{\hskip .11em plus .33em minus -.07em}
 \sloppy
 \sfcode`\.=1000\relax}

\begin{document}
\preprint{}
\title{%
Supergravity inspired Vector Curvaton
}

\author{Konstantinos Dimopoulos}%
\email{k.dimopoulos1@lancaster.ac.uk}
\affiliation{%
Physics Department, Lancaster University\\}%

\date{\today}

\begin{abstract}
It is investigated whether a massive Abelian vector field, whose gauge kinetic 
function is growing during inflation, can be responsible for the generation of 
the curvature perturbation in the Universe. Particle production is studied and 
it is shown that the vector field can obtain a scale invariant superhorizon 
spectrum of perturbations with a reasonable choice of kinetic function.
After inflation the vector field begins coherent oscillations, during which it
corresponds to pressureless isotropic matter. When the vector field dominates 
the Universe its perturbations give rise to the observed curvature perturbation
following the curvaton scenario. It is found that this is possible if, after 
the end of inflation, the mass of the vector field increases
at a phase transition at temperature of order 1~TeV or lower.
Inhomogeneous reheating, whereby the vector field modulates the decay rate of 
the inflaton, is also studied.
\end{abstract}

\maketitle

\section{Introduction}
Observations suggest that the formation of structure in the Universe is
due to the existence of a primordial spectrum of superhorizon curvature 
perturbations. The fact that they are superhorizon strongly supports the
hypothesis that these perturbations were generated through an acausal 
process. The most compelling mechanism for this to date is cosmic inflation.

During inflation, all light, non-conformally invariant fields undergo particle 
production and obtain a superhorizon spectrum of perturbations. These 
perturbations can be responsible for the generation of the curvature
perturbation in the Universe, if their spectrum is compatible with the 
observations. Traditionally, it has been considered that it is the 
perturbations of the inflaton field itself, which give rise to the curvature 
perturbation. However, this inflaton hypothesis typically results in
overconstraining inflation model-building, which leads to fine-tuning. 

For this reason, alternative suggestions have been recently put forward. 
According to such proposals, the field responsible for the curvature 
perturbation may have little or nothing to do with the dynamics of inflation. 
One possibility is to consider a field whose contribution to the density is 
negligible during inflation but, after the end of inflation, it manages to 
dominate (or nearly dominate) the Universe before its decay, thereby imposing 
its own curvature perturbation spectrum. This is the so-called curvaton 
hypothesis \cite{curv}. Under this hypothesis the fine-tunning problems of 
inflation are alleviated \cite{liber,liber2}, while one can attain inflation 
at low energy scales \cite{low,low1,laza}. Many suggestions in the literature 
offer realistic candidates in theories beyond the standard model, which can 
play the role of the curvaton field. 

Another suggestion along similar lines is that the field responsible for the 
curvature perturbation is not related to the dynamics of inflation but it 
affects the reheating process by modulating the decay rate of the inflaton. 
This is the inhomogeneous reheating mechanism \cite{inhom,inhom2}, which can 
also allow for low-scale inflation \cite{inhomlow}.

Until now the literature considers that the curvature perturbation in the 
Universe is due to particle production of a suitable {\em scalar field}, 
typically through one of the above mechanisms. However, even though theories 
beyond the standard model (in particular supersymmetric theories) contain a 
plethora of scalar fields, the fact that no scalar field has been observed as 
yet undermines somewhat the predictability and falsifiability of these models. 
In contrast, in this paper, we consider the possibility that the curvature
perturbation is due to particle production of a {\em vector field} during 
inflation.

A massive vector field is non-conformally invariant and can indeed undergo 
particle production during inflation. In Ref.~\cite{vec} this scenario has been
investigated for a massive Abelian vector field. It was shown that a 
scale-invariant spectrum of perturbations can be generated provided the mass of
the vector field satisfies the condition \mbox{$m^2\approx -2H_*^2$} during 
inflation, where $H_*$ is the inflationary Hubble scale. However, this 
condition is hard to realise in a theoretically well motivated way. 

This problem is overcome in this paper by considering a non-trivial evolution 
of the kinetic term for the vector field, during inflation. In supergravity the
kinetic term of vector fields is determined, in general, by the gauge kinetic 
function which is a holomorphic function of the fields of the theory. We 
consider a similar setup here and assume that, the kinetic function is 
dominated by a degree of freedom which varies substantially during inflation,
while the cosmological scales exit the horizon. We find that a scale-invariant
spectrum of vector field perturbations can be attained, without the need for a 
negative mass-squared for the vector field, if the kinetic function is growing 
with time during inflation.

We then investigate how such a spectrum of vector field perturbations can give
rise to the observed curvature perturbation in the Universe. In general, a 
homogeneous vector field generates an anisotropic pressure, which, if dominant,
results in a large-scale anisotropy that contradicts the observations (isotropy
of the CMB). This is why, the vector field cannot play the role of the 
inflaton (see, however, \cite{VI}). On the other hand, as shown in 
Ref.~\cite{vec}, a massive oscillating vector field has zero average pressure 
and behaves as pressureless, isotropic matter. 
Thus, it can safely dominate the Universe without generating a long-range 
anisotropy. Hence, one can employ the curvaton mechanism to generate the 
curvature perturbation in the Universe, using as curvaton a massive vector 
field, which has assumed a scale-invariant spectrum of perturbations during 
inflation. In this paper we study in detail the use of such a vector field as 
curvaton.

One other way to attempt to generate the curvature perturbation from the 
vector field without the latter ever dominating the Universe, is by 
considering that the vector field controls the decay rate of the inflaton,
resulting in inhomogeneous reheating. We briefly investigate this scenario as 
well.

The paper is structured as follows. In Sec.~II we derive the equations of 
motion for the perturbations of a massive vector field with a varying kinetic 
function and mass. In Sec.~III we study particle production during inflation
of this vector field and obtain the necessary conditions to attain the desired
scale-invariant spectrum. In Sec.~IV we study the dynamics of the scalar field
which controls the kinetic function for our vector field. In Sec.~V we obtain
the spectrum of the produced perturbations in the case when the vector field 
has constant mass and also when its mass is controlled by the scalar field 
which also controls the kinetic function. In Sec.~VI we study analytically the 
curvaton scenario. By obtaining the energy-momentum tensor for the vector field
we find the scaling of its density during and after inflation and reheating.
We then implement this to find the parameter space in which the vector field 
can generate the curvature perturbation. We find that the lower bound on the 
inflationary scale is too stringent to allow the scenario to work. In Sec.~VII 
we employ the mass increment mechanism to lower further the inflationary scale.
The mechanism assumes that the mass of the vector field grows at a phase 
transition
after the end of inflation. In Sec.~VIII we study possible complications to the
scenario due to the dynamics and to particle production of the scalar field 
that controls the kinetic function. In Sec.~IX we present a concrete example 
of our vector curvaton model, taking all the constraints into account. In
Sec.~X the inhomogeneous reheating mechanism is tried out, using as the 
inflaton the field that controls the kinetic function. The mechanism is shown 
to be ineffective. Finally, in Sec.~XI we discuss our results and present our
conclusions.

Throughout the paper we use natural units, where
\mbox{$c=\hbar=1$} and Newton's gravitational constant is
\mbox{$8\pi G=m_P^{-2}$}, with \mbox{$m_P=2.4\times 10^{18}$GeV} being the
reduced Planck mass. The signature of the metric is \mbox{(1,-1,-1,-1)}.

\section{The equations of motion}\label{teom}

Consider the following Lagrangian density for a massive vector field with 
mass $m$
\begin{equation}
{\cal L}=-\frac{1}{4}fF_{\rm \mu\nu}F^{\mu\nu}+\frac{1}{2}m^2A_\mu A^\mu\;,
\label{L}
\end{equation}
where $f=f(t)$ is a function of cosmic time $t$ reminiscent of the gauge 
kinetic function in supergravity.\footnote{A similar setup is employed in 
so-called dilaton electromagnetism \cite{bambayoko}, where 
\mbox{$f=e^{-\lambda\Phi/m_P}$} with $\Phi$ being the dilaton. This setup has 
been used to break the conformality of electromagnetism and generate a 
primordial magnetic field during inflation \cite{bambasasa} (see 
also~\cite{ratra}).}
In general, the mass of the vector field can 
also depend on time, i.e. \mbox{$m=m(t)$}.
For an Abelian field, the field strength tensor is
\begin{equation}
F_{\mu\nu}=
\partial_\mu A_\nu - \partial_\nu A_\mu
\;.
\label{F}
\end{equation}
Employing the above one obtains the field equations for the vector field:
\begin{equation}
\,\!
\left[\partial_\mu+
\left(\partial_\mu\ln\sqrt{-{\cal G}}\right)\right]
[f(\partial^\mu A^\nu-\partial^\nu A^\mu)]+m^2A^\nu=0\,,
\hspace{-.5cm}
\label{FE}
\end{equation}
where \mbox{$\cal G$} is the determinant of the metric tensor.

Since we are interested in particle production during inflation we assume
that, to a good approximation, the spacetime is spatially flat, homogeneous and
isotropic. Hence, we use the flat-FRW metric:
\begin{equation}
ds^2=dt^2-a^2(t)dx^idx^i,
\label{FRW}
\end{equation}
where $a=a(t)$ is the scale factor of the Universe, $x^i$ are Cartesian 
spatial coordinates with $i=1,2,3$ and Einstein summation is assumed. 
Employing the above metric into Eq.~(\ref{FE}) and following the process 
detailed in Ref.~\cite{vec} we obtain the following temporal and spatial 
components of the field equations respectively:
\begin{equation}
\mbox{\boldmath $\nabla\cdot\dot{A}$}
-\nabla^2 A_t+\frac{(am)^2}{f} A_t=0\,,
\label{v=0}
\end{equation}
and
\begin{eqnarray}
& & \mbox{\boldmath $\ddot A$}+\left(H+\frac{\dot f}{f}\right)
\mbox{\boldmath $\dot A$}+\frac{m^2}{f}\mbox{\boldmath $A$}
-a^{-2}\nabla^2\mbox{\boldmath $A$}\;=\nonumber\\
 & & =\;\left(\frac{\dot f}{f}-2\frac{\dot m}{m}-2H\right)
\mbox{\boldmath $\nabla$}A_t\;,
\label{EoM}
\end{eqnarray}
where the dot denotes derivative with respect to the cosmic time and
\mbox{\boldmath $\nabla$} stands for the divergence or the gradient
while \mbox{$\nabla^2\equiv\partial_i\partial_i$} is the Laplacian.

We expect inflation to homogenise the vector field and, therefore,
\begin{equation}
\partial_i A_\mu=0\quad\forall\quad\mu\in[0,3]\,.
\label{hom0}
\end{equation}
Enforcing this condition into Eq.~(\ref{v=0}) we obtain 
\begin{equation}
A_t=0\,.
\label{At=0}
\end{equation}
Using Eqs.~(\ref{hom0}) and (\ref{At=0}) into Eq.~(\ref{EoM}) we find
\begin{equation}
\mbox{\boldmath $\ddot A$}+\left(H+\frac{\dot f}{f}\right)
\mbox{\boldmath $\dot A$}+\frac{m^2}{f}\mbox{\boldmath $A$}=0\,.
\label{EoMhom}
\end{equation}
The above is reminiscent of the Klein-Gordon equation of a homogeneous scalar 
field in an expanding Universe, with the crucial difference that the 
coefficient in the ``friction'' term does not feature a factor of $3H$.

We are interested in the generation of superhorizon perturbations of the
vector field, which might be responsible for the curvature perturbations in
the Universe. Therefore, we perturb the vector field around the homogeneous
value $A_\mu(t)$ as follows:
\begin{equation}
\begin{array}{l}
A_\mu(t, \mbox{\boldmath $x$})=A_\mu(t)+\delta A_\mu(t, \mbox{\boldmath $x$})
\quad\Rightarrow\\
\\
\mbox{\boldmath $A$}(t, \mbox{\boldmath $x$})=
\mbox{\boldmath $A$}(t)+\delta \mbox{\boldmath $A$}(t, \mbox{\boldmath $x$})
\;\;\&\;\;
A_t(t, \mbox{\boldmath $x$})=\delta A_t(t, \mbox{\boldmath $x$}),\!\!\!\!\!\!
\end{array}
\label{pert}
\end{equation}
where we took into account Eq.~(\ref{At=0}). In the above 
\mbox{\boldmath $A$}$(t)$ satisfies Eq.~(\ref{EoMhom}). In view of 
Eqs.~(\ref{EoMhom}) and (\ref{pert}), Eqs.~(\ref{v=0}) and (\ref{EoM}) become
\begin{eqnarray}
\hspace{-1cm} & &
\mbox{\boldmath $\nabla\cdot$}\dot{(\delta\mbox{\boldmath $A$})}
-\nabla^2 \delta A_t+\frac{(am)^2}{f} \delta A_t=0
\label{v=0pert}\\
\hspace{-1cm} & & \nonumber\\
\hspace{-1cm} & &
\ddot{(\delta\mbox{\boldmath $A$})}+
\left(H+\frac{\dot f}{f}\right)\dot{(\delta\mbox{\boldmath $A$})}+
\frac{m^2}{f}\delta\mbox{\boldmath $A$}
-a^{-2}\nabla^2\delta\mbox{\boldmath $A$}=\nonumber\\
\hspace{-1cm} & &
=\;\left(\frac{\dot f}{f}-2\frac{\dot m}{m}-2H\right)
\mbox{\boldmath $\nabla$}\delta A_t\,.
\label{EoMpert}
\end{eqnarray}

Now, let us switch to momentum space by Fourier expanding the perturbations:
\begin{equation}
\delta A_\mu(t, \mbox{\boldmath $x$})=
\int\frac{d^3k}{(2\pi)^{3/2}}\;\delta{\cal A}_\mu (t, \mbox{\boldmath $k$})\,
\exp(i\mbox{\boldmath $k\cdot x$})\,.
\label{fourier}
\end{equation}
Using the above, Eq.~(\ref{v=0pert}) becomes
\begin{equation}
\delta{\cal A}_t+
\frac{i\partial_t(\mbox{\boldmath $k\cdot$}\delta\mbox{\boldmath $\cal A$})}%
{k^2+(am)^2/f}=0\,,
\label{calAt}
\end{equation}
where \mbox{$k^2\equiv$}~\mbox{\boldmath $k\cdot k$}. Using this and 
Eq.~(\ref{fourier}) we can write Eq.~(\ref{EoMpert}) as
\begin{eqnarray}
& & \ddot{(\delta\mbox{\boldmath $\cal A$})}+
\left(H+\frac{\dot f}{f}\right)\dot{(\delta\mbox{\boldmath $\cal A$})}+
\frac{m^2}{f}\delta\mbox{\boldmath $\cal A$}+
\left(\frac{k}{a}\right)^2\!\delta\mbox{\boldmath $\cal A$}+
\nonumber\\
& & +\;
\left(2H+2\frac{\dot m}{m}-\frac{\dot f}{f}\right)\frac{\mbox{\boldmath $k$}
\partial_t(\mbox{\boldmath $k\cdot$}\delta\mbox{\boldmath $\cal A$})}%
{k^2+(am)^2/f}=0\,.
\label{calEoM}
\end{eqnarray}

We can rewrite the above in terms of the components parallel and perpendicular
to \mbox{\boldmath $k$}, defined as:
\begin{equation}
\delta\mbox{\boldmath $\cal A$}^\parallel\equiv
\frac{\mbox{\boldmath $k$}(\mbox{\boldmath $k\cdot$}%
\delta\mbox{\boldmath $\cal A$})}{k^2}
\quad\&\quad
\delta\mbox{\boldmath $\cal A$}^\perp\equiv
\delta\mbox{\boldmath $\cal A$}-\delta\mbox{\boldmath $\cal A$}^\parallel.
\label{perpparal}
\end{equation}
Thus, we obtain the following equations of motion for 
the vector field perturbations in momentum space:
\begin{eqnarray}
& & \hspace{-1cm} 
\left[\partial_t^2+\left(H+\frac{\dot f}{f}\right)\partial_t+
\frac{m^2}{f}+\left(\frac{k}{a}\right)^2\right]
\delta\mbox{\boldmath $\cal A$}^\perp 
=0
\label{EoMperp}\\
 & & \nonumber\\
& & \hspace{-1cm} \left\{\partial_t^2+
\left[H+\frac{\dot f}{f}+
\frac{\left(2H+2\frac{\dot m}{m}-\frac{\dot f}{f}\right)k^2}{k^2+
\frac{(am)^2}{f}}\right]
\partial_t\,+\right. \nonumber\\
& &  \hspace{-1cm} \left.+\;\frac{m^2}{f}+\left(\frac{k}{a}\right)^2\right\}
\delta\mbox{\boldmath $\cal A$}^\parallel=0\,.
\label{EoMparal}
\end{eqnarray}

\section{Particle production}\label{pp}

To investigate particle production during inflation for the vector field
we need to solve the equation of motion for the perturbations of the field.
The integration constants are then evaluated by matching the solution to
the vacuum at early times (when \mbox{$k/aH\rightarrow+\infty$}), i.e. by
demanding
\begin{equation}
\lim_{_{\hspace{.5cm}
\frac{k}{aH}\rightarrow+\infty}}\hspace{-.5cm}
\delta{\cal A}_k=
\frac{1}{\sqrt{2k}}\exp(ik/aH),
\label{match}
\end{equation}
where \mbox{$\delta{\cal A}_k\equiv\delta$\mbox{\boldmath $\cal A$}
$(t, \mbox{\boldmath $k$})$} and we note that at early times the perturbation 
in question is well within the horizon, which means that 
\mbox{$a\rightarrow 1$} and \mbox{$k/aH\rightarrow kt$}.

Afterwards we evaluate the solution at late times, when the perturbation is 
superhorizon in size (i.e. when \mbox{$k/aH\rightarrow 0^+$}). The power 
spectrum is obtained by 
\begin{equation}
{\cal P_A}=\frac{k^3}{2\pi^2}\left|\hspace{-.5cm}\lim_{_{\hspace{.5cm}
\frac{k}{aH}\rightarrow\mbox{\scriptsize 0}^{^{+}}}}\hspace{-.5cm}
\delta{\cal A}_k\right|^2.
\label{PA}
\end{equation}

We assume that, during inflation, $H$ is constant. We also assume that
$f$ is proportional to some power of the scale factor, such that
\begin{equation}
f\propto a^{\alpha-1}\;\Rightarrow\frac{\dot f}{f}=(\alpha-1)H\,,
\label{fa}
\end{equation}
with $\alpha$ being a constant. 

We will concern ourselves only with the transverse component of the vector 
field perturbations Eq.~(\ref{EoMperp}), whose equation of motion we write as
\begin{equation}
\left[\partial_t^2+\alpha H\partial_t+
\tilde m^2+\left(\frac{k}{a}\right)^2\right]
\delta{\cal A}_k =0\,,
\label{EoMfin}
\end{equation}
where $\tilde m$ is a constant associated with the mass $m$ of the vector 
field (see below) and we have dropped the `$\perp$' for simplicity.

Solving Eq.~(\ref{EoMfin}) and matching to the vacuum in Eq.~(\ref{match}), 
we obtain the solution
\begin{equation}
\delta{\cal A}_k
=\frac{1}{2}\sqrt{\frac{\pi}{aH}}
e^{i(\nu+\frac{1}{2})\frac{\pi}{2}}
H_\nu^{(1)}(k/aH)\,,
\label{solu}
\end{equation}
where with $H_\nu^{(1)}$ we denote the Hankel function of the first kind and
\begin{equation}
\nu\equiv
\sqrt{\left(\frac{\alpha}{2}\right)^2-\left(\frac{\tilde m}{H}\right)^2}.
\label{v}
\end{equation}
The above solution at late times approaches
\begin{eqnarray}
 & & \hspace{-1.2cm}
\lim_{_{\hspace{.5cm}
\frac{k}{aH}\rightarrow\mbox{\scriptsize 0}^{^{+}}}}\hspace{-.5cm}
\delta{\cal A}_k =
\frac{1}{2}\sqrt{\frac{\pi}{aH}}
e^{i(\nu+\frac{1}{2})\frac{\pi}{2}}
\;\times\nonumber\\
 & & \nonumber\\
 & & \hspace{-1.2cm}\times
\left[\frac{1+i\pi(\frac{1}{2}-\nu)}{\Gamma(1+\nu)}
\left(\frac{k}{2aH}\right)^\nu\!\!-
\frac{i}{\Gamma(1-\nu)}\left(\frac{k}{2aH}\right)^{-\nu}
\right]\!.
\label{late1}
\end{eqnarray}
Hence, using Eq.~(\ref{PA}) we find that the dominant contribution to the 
power spectrum is
\begin{equation}
{\cal P_A}\approx\frac{4\pi}{|\Gamma(1-\nu)|^2}
\left(\frac{aH}{2\pi}\right)^2\left(\frac{k}{2aH}\right)^{3-2\nu}.
\label{PAperp}
\end{equation}

Therefore, we may obtain a scale-invariant spectrum if
\begin{equation}
\nu\approx 3/2\quad\Leftrightarrow\quad
\left(\frac{\alpha}{2}\right)^2\approx\frac{9}{4}+
\left(\frac{\tilde m}{H}\right)^2.
\label{v=3/2}
\end{equation}
In this case we find that a scale invariant spectrum of perturbations is 
recovered with
\begin{equation}
{\cal P_A}\approx a^2\left(\frac{H}{2\pi}\right)^2.
\label{PAflat}
\end{equation}
as in the case of a massless scalar field. 

Parameterising the scale 
dependence of the perturbations in the usual manner
\begin{equation}
{\cal P_A}(k)\propto k^{n_s-1},
\label{PAns}
\end{equation}
and comparing with Eq.~(\ref{PAperp}) we obtain, for the spectral index, the 
result
\begin{equation}
n_s-1=3-2\nu=3-\alpha\sqrt{1-\left(\frac{2\tilde m}{\alpha H}\right)^2}\,,
\label{ns}
\end{equation}
where we also used Eq.~(\ref{v}). 
In the case when \mbox{$\tilde m\ll H$} we find
\begin{equation}
n_s\simeq(4-\alpha)-\frac{6}{\alpha}\eta
\quad{\rm where}\quad \eta\equiv\frac{1}{3}\left(\frac{\tilde m}{H}\right)^2,
\label{nseta}
\end{equation}
which, when $\alpha=3$, is the usual finding in the case of a light scalar 
field.\footnote{There is no contribution from 
\mbox{$\epsilon\equiv-\dot H/H^2$} to the spectral 
index because we have taken \mbox{$H=$ const}.}

\section{Fast-Rolling Scalar Field}\label{frsf}

From Eq.~(\ref{fa}) we see that, if $f=$~constant then $\alpha=1$ and we can
have a scale invariant spectrum of perturbations only if 
\mbox{$\tilde m^2\approx -2H^2$} (c.f Eq.~(\ref{v=3/2})), i.e. only if the 
effective mass-squared of the vector field is negative \cite{vec}. 
To avoid this, we need to consider that $f(t)$ is
controlled by a degree of freedom which undergoes non-trivial evolution during 
inflation, at least during the period when the cosmological scales exit the 
horizon. This is natural to expect in supergravity.

Indeed, in supergravity, the scalar fields of the theory receive a contribution
to their mass of order the Hubble scale $H$ during inflation, due to 
corrections to the scalar potential generated by a non-minimal K\"{a}hler 
potential \cite{eta}. Hence, these scalar fields are expected to evolve 
substantially during inflation as they fast-roll down the potential slopes. 
Hence, dependence of $f$ on these scalar fields is expected to yield naturally 
\mbox{$\dot f\neq 0$} during inflation. To parametrise this behaviour we assume
that $f$ is a function of some scalar field $\phi=\phi(t)$, whose value varies 
during inflation. 

The gauge kinetic function in supergravity is a holomorphic function of the 
fields of the theory. Hence, we consider that $f(\phi)$ can be expanded around
the origin as \mbox{$f(\phi)\approx\sum_{1/2}^\infty c_n(\phi/M)^{2n}$}, where 
$M$ is some cutoff scale and $c_n$ are constant 
coefficients. We assume that this sum is dominated by a term of $n$-th order, 
so that we can write
\begin{equation}
f(\phi)\simeq\left(\frac{\phi}{M}\right)^{2n},
\label{fn}
\end{equation}
where we have absorbed $c_n$ into $M$. 
Inserting the above into Eq.~(\ref{fa}) we find
\begin{equation}
\phi\propto a^{\frac{\alpha-1}{2n}}.
\label{phia}
\end{equation}

Let us introduce the following Lagrangian density for the scalar field $\phi$:
\begin{equation}
{\cal L}_\phi=\frac{1}{2}\partial_\mu\phi\partial^\mu\phi-V(\phi)\,,
\label{Lphi1}
\end{equation}
where, for the scalar potential, we consider
\begin{equation}
V(\phi)=V_0-\frac{1}{2}m_\phi^2\phi^2+\cdots,
\label{Vphi}
\end{equation}
where the dots denote higher order terms which stabilise the potential at
\mbox{$\phi_{\rm vev}=M$}, such that
\begin{equation}
V_0\sim m_\phi^2M^2.
\label{V0}
\end{equation}
Hence, the kinetic term of the vector field becomes canonical ($f=1$) after
$\phi$ settles at its vacuum expectation value (VEV).

Assuming that the field has been homogenised by inflation, its equation of 
motion, when \mbox{$\phi<M$}, is
\begin{equation}
\ddot\phi+3H\dot\phi-m_\phi^2\phi\simeq 0\,.
\label{EoMphi}
\end{equation}
The solution of the above during inflation has a growing mode of the form
\begin{equation}
\phi\simeq\phi_0\exp\left\{\frac{3}{2}H\Delta t
\left[\sqrt{1+\frac{4}{9}\left(\frac{m_\phi}{H}\right)^2}-1\right]\right\},
\label{phiDt}
\end{equation}
where $\phi_0$ is the initial value at the onset of inflation and $\Delta t$
is the elapsed time. Comparing the above with Eq.~(\ref{phia}) we find
\begin{equation}
\frac{m_\phi}{H}=\frac{3}{2}\sqrt{\left(\frac{\alpha-1}{3n}+1\right)^2-1}\,.
\label{mphiH}
\end{equation}
where we considered that $a\propto e^{H\Delta t}$. The above means that, if
\mbox{$\alpha,n={\cal O}(1)$} then \mbox{$m_\phi\sim H$} during inflation.
This is naturally expected for scalar fields in supergravity due to corrections
introduced to the scalar potential when considering a generic form of the 
K\"{a}hler potential \cite{eta}. This is the source of the so-called 
$\eta$-problem, which is endemic to inflation when a scalar field is used to 
produce the curvature perturbation in the Universe.

From Eq.~(\ref{phia}) it is easy to obtain the number of e-folds it takes for
$\phi$ to reach the minimum of $V(\phi)$:
\begin{equation}
N_\phi=\frac{2n}{\alpha-1}\ln\left(\frac{M}{\phi_0}\right).
\label{Nphi}
\end{equation}
After inflation \mbox{$H(t)<m_\phi$}, which means that $\phi$ rushes toward its
VEV, \mbox{$\phi_{\rm vev}=M$}, in less than a Hubble time.

\section{Spectrum of perturbations}\label{sp}

In this section we concentrate on two particular possibilities, which may be
realised in this model. Other possibilities exist but the following appear to
be the most straightforward for investigation.

\subsection{Constant mass}

Suppose at first that the mass of the vector field is constant. In this case,
the mass term in Eq.~(\ref{EoMperp}) is
\begin{equation}
\frac{m^2}{f}\propto a^{1-\alpha}.
\end{equation}

Let us choose $\alpha=3$.\footnote{Another interesting choice is 
\mbox{$\alpha=1$} because, in this case \mbox{$m^2/f=$~const.} However, by
vitrue of Eq.~(\ref{phia}), such a choice implies that \mbox{$\phi=$~const.}
which means that \mbox{$f=$~const.} (c.f. Eq.~(\ref{fn})). This case, 
therefore, is already explored in Ref.~\cite{vec} and requires a negative 
mass-squared for the vector field.}
Then the above suggests that \mbox{$m^2/f\propto a^{-2}$},
which means that the mass term in Eq.~(\ref{EoMperp}) scales as the $(k/a)^2$ 
term. Thus, the resulting equation of motion is of the form of 
Eq.~(\ref{EoMfin}) with $\alpha=3$ and $\tilde m=0$ under the substitution:
\begin{equation}
k\rightarrow k'\quad{\rm where}\quad
k'\equiv\sqrt{k^2+k_c^2}\,,
\end{equation}
where 
\begin{equation}
k_c^2\equiv\frac{(am)^2}{f}=(a_0m)^2\left(\frac{M}{\phi_0}\right)^{2n}
=(a_0m)^2e^{2N_\phi},
\label{kc}
\end{equation}
with $a_0$ being the value of the scale factor at the onset of inflation
and we have used Eq.~(\ref{Nphi}).
The solution of Eq.~(\ref{EoMperp}) is, therefore, the one described in 
Eq.~(\ref{solu}), which, at late times, approaches the result in 
Eq.~(\ref{late1}), with \mbox{$k\rightarrow k'$} and $\nu=3/2$. 
Hence, in view of Eq.~(\ref{PA}) we obtain the dominant contribution to the
power spectrum:
\begin{equation}
{\cal P_A}\approx\left(\frac{aH}{2\pi}\right)^2
\left(\frac{k^2}{k^2+k_c^2}\right)^{3/2}.
\label{PAm}
\end{equation}

Thus, when \mbox{$k\gg k_c$}, the power spectrum is approximately scale 
invariant. In the opposite case, \mbox{${\cal P_A}\propto k^3$}. If these
perturbations are to give rise to the curvature perturbations in the Universe
the cosmological scales should correspond to scales with \mbox{$k_*>k_c$}.
Hence, we require:
\begin{eqnarray}
& & \frac{k_c}{H}=a_c<a_*\equiv a_{\rm end}e^{-N_*}\nonumber\\
& \Rightarrow & \frac{m}{H}<\exp(N_{\rm tot}-N_*-N_\phi)\,,
\label{mcons}
\end{eqnarray}
where $a_c$ is the scale factor at the time when the scale $k_c$ exits the 
horizon during inflation, $a_{\rm end}$ is the scale factor at the end of 
inflation, the subscript `*' denotes the time when the cosmological scales 
exit the horizon, \mbox{$N_{\rm tot}\equiv a_{\rm end}/a_0$} denotes the
total number of e-folds of inflation and we have used Eq.~(\ref{kc}).

The condition in Eq.~(\ref{mcons}) can be better understood when considering 
the ``effective'' mass of the vector field during inflation as featured in the
equation of motion (\ref{EoMhom}):
\begin{eqnarray}
& & \frac{m^2}{f}=m^2e^{2N_\phi}\left(\frac{a_0}{a}\right)^2
\;\Rightarrow\nonumber\\
& \Rightarrow & \frac{m}{\sqrt f}=m\exp(N_\phi+N-N_{\rm tot})\,,
\label{meff}
\end{eqnarray}
where we have used Eqs.~(\ref{kc}) and (\ref{Nphi}) with 
\mbox{$\alpha=3$}. In view of the above we see that the constraint in 
Eq.~(\ref{mcons}) corresponds to the requirement:
\begin{equation}
\left.\frac{m^2}{f}\right|_*<H\,.
\label{mlight}
\end{equation}

If $\phi$ were also responsible for inflation we would have 
\mbox{$N_{\rm tot}=N_\phi$} and the above constraint would reed 
\mbox{$m<e^{-N_*}H$}. According to Ref.~\cite{vec}, satisfying this bound 
allows the generation of a scale-invariant perturbation spectrum for the
longitudinal component of the vector field, which may also be used to
generate the curvature perturbation in the Universe. However, as discussed in 
Ref.~\cite{vec}, such a bound on the mass of the vector field is very hard to 
satisfy. Hence, most probably, $\phi$ needs to be some scalar field other than 
the inflaton
(see also footnote~\ref{MmP}). In this case too, 
though, we need inflation not to last too long because the cosmological scales 
have to exit the horizon while $\phi$ is still rolling. Otherwise, the roll of 
$\phi$ is irrelevant and we are back to the case studied in Ref.~\cite{vec}. 

\begin{figure}[t]
\includegraphics[width=80mm,angle=0]{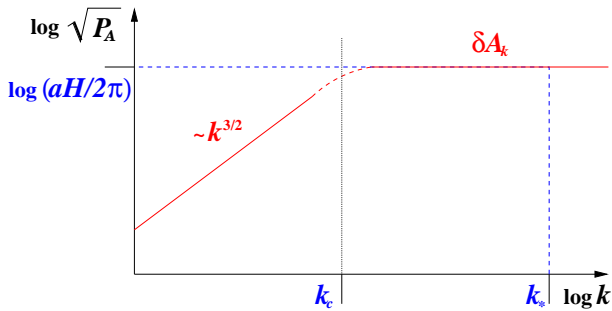}
\caption{
Illustration of the superhorizon spectrum of the transverse 
component of the perturbation of a sufficiently light vector field
in the case when \mbox{$m=$~const.} and \mbox{$f\propto a^2$}. 
At momenta smaller than $k_c$ the spectrum is 
\mbox{$\delta A_k\propto k^{3/2}$}, while when \mbox{$k\gg k_c$}
the spectrum becomes approximately scale invariant
\mbox{$\delta A_k\approx aH/2\pi$}. Hence, the cosmological scales should
correspond to momentum $k_*>k_c$.}
\end{figure}

\subsection{Higgsed vector field}\label{hvf}

Suppose, now that the mass of the vector field is due to an interaction between
the former and the scalar field $\phi$. In this case, the Lagrangian of the
model is
\begin{equation}
{\cal L}=-\frac{1}{4}f(\phi)F_{\rm \mu\nu}F^{\mu\nu}+
\frac{1}{2}D_\mu\phi(D^\mu\phi)^*-V(\phi)\,,
\label{L1}
\end{equation}
where $V(\phi)$ is given by Eq.~(\ref{Vphi}),
\mbox{$D_\mu\phi=\partial_\mu\phi-igA_\mu\phi$} is the covariant 
derivative in field space and $g$ is the (gauge) coupling 
($\phi$ is taken to be real for simplicity). Then, the mass term of the vector
field is
\begin{equation}
\delta{\cal L}=\frac{1}{2}g^2\phi^2A_\mu A^\mu\;,
\label{dL}
\end{equation}
i.e. the mass of the vector field, in this case, is \mbox{$m=g\phi$}.
Consequently, this time, the mass term in Eq.~(\ref{EoMperp}) is
\begin{equation}
\frac{m^2}{f}\propto\phi^{2(1-n)}\propto a^{(\frac{n-1}{n})(1-\alpha)},
\label{mf1}
\end{equation}
where we used Eqs.~(\ref{fn}) and (\ref{phia}).

Now, one might be interested to obtain a scale invariant spectrum in the
same manner as the previous subsection, i.e. by taking 
\mbox{$m^2/f\propto a^{-2}$}. As we have seen, this case corresponds to
\mbox{$\tilde m=0$} in Eq.~(\ref{EoMfin}). Then, according to 
Eq.~(\ref{v=3/2}), scale invariance requires \mbox{$\alpha=3$}. However,
in view of Eq.~(\ref{mf1}), this is only possible when 
\mbox{$n\gg 1$}, which is not realistic. Thus, it seems that 
\mbox{$m^2/f\propto a^{-2}$} is not realisable in this case.

Another option is to consider that \mbox{$m^2/f=$ const}. From Eq.~(\ref{mf1})
we see that this is possible if either $\alpha=1$ or $n=1$. The former case
implies that \mbox{$\phi=$ const.} (cf. Eq.~(\ref{phia})), which means that
\mbox{$f=$ const.} (cf. Eq.~(\ref{fn})). This case has been explored in 
Ref.~\cite{vec} and requires a negative mass-squared for the vector field. 
Let us then concentrate in the latter case, when \mbox{$n=1$}.

Assuming $n=1$ means that Eq.~(\ref{EoMhom}) becomes
\begin{equation}
\mbox{\boldmath $\ddot A$}+\alpha H\mbox{\boldmath $\dot A$}+
(gM)^2\mbox{\boldmath $A$}=0\,,
\label{eom}
\end{equation}
where we have used Eq.~(\ref{fa}) and that $m=g\phi$ with
\begin{equation}
f=\frac{\phi^2}{M^2}.
\label{f2}
\end{equation}
Therefore, the mass term in Eq.~(\ref{EoMperp}) becomes 
\begin{equation}
\frac{m^2}{f}=(gM)^2,
\label{meff1}
\end{equation}
which suggests that \mbox{$\tilde m=gM$} in this case. Hence, according to 
Eq.~(\ref{v=3/2}), scale invariance requires
\begin{equation}
\left(\frac{\alpha}{2}\right)^2\approx\frac{9}{4}+
\left(\frac{gM}{H}\right)^2,
\label{scinv}
\end{equation}
which means that \mbox{$\alpha\geq 3$}.

In view of the above, the solution to Eq.~(\ref{eom}), during inflation, is
\begin{equation}
|\mbox{\boldmath $A$}|=
C_1
e^{-\frac{1}{2}(\alpha+3)H\Delta t}+
C_2e^{\frac{1}{2}(3-\alpha)H\Delta t}
\propto a^{\frac{3-\alpha}{2}},
\label{solu1}
\end{equation}
where 
$C_1$ and $C_2$
are constants of integration and, in the proportionality relation, we have 
considered only the ``growing mode''. Hence we see that,
for \mbox{$\alpha>3$}, the magnitude of the vector field is decreasing. This is
undesirable, as will be made clear later (see footnote~\ref{foot}). 
Hence, we choose \mbox{$\alpha=3$},
in which case $|\mbox{\boldmath $A$}|\simeq$~const. To satisfy, therefore, 
Eq.~(\ref{scinv}), we need to enforce the constraint
\begin{equation}
gM<\frac{3}{2}H\,.
\label{massless0}
\end{equation}

The above constraint is necessary in order to obtain an approximately scale 
invariant spectrum of perturbations. However, if these perturbations are to 
account for the curvature perturbations in the Universe then the above 
constraint is tightened further by spectral index considerations. Indeed, from 
Eq.~(\ref{ns}) we readily find
\begin{equation}
n_s-1\simeq\frac{2}{3}\left(\frac{gM}{H}\right)^2.
\label{nshiggs}
\end{equation}
Hence, the spectrum obtained is blue in contrast to the observational 
preferences. Since $n_s\approx 1.00$ is still marginally acceptable and the
precision of the observational data is at the level of a few percent, we 
obtain the following bound
\begin{equation}
gM\lsim 0.1\,H\,.
\label{massless}
\end{equation}

\section{Vector Curvaton}

One mechanism for generating the curvature perturbation in the Universe
starting from a superhorizon spectrum of vector field perturbations, follows
the curvaton scenario. In this case, the vector field, while subdominant 
during inflation, may come to dominate (or nearly dominate) some time 
afterwards. When it does so, it imposes its own curvature perturbation onto the
Universe~\cite{curv}. 

\subsection{The energy momentum tensor}\label{emt}

To compute if and when the vector field dominates the Universe, in order
to imprint its superhorizon perturbation spectrum, we follow the evolution of 
the energy-momentum tensor of the vector field.

Using Eq.~(\ref{L}), the energy momentum tensor for $A_\mu$ is
\begin{eqnarray}
T_{\mu\nu} & = & 
f\left(\frac{1}{4}g_{\mu\nu}F_{\rho\sigma}F^{\rho\sigma}-
F_{\mu\rho}F_\nu^{\;\rho}\right)+ \nonumber\\
& &
+\,m^2\left(A_\mu A_\nu-\frac{1}{2}g_{\mu\nu}A_\rho A^\rho\right).
\label{Tmn}
\end{eqnarray}

Assume that the homogenised vector field lies along the $z$-direction
\begin{equation}
A_\mu=(0,0,0,A(t)\,)\,.
\label{A}
\end{equation}
Then the energy-momentum tensor can be written in the form
\begin{equation}
T_\mu^{\,\nu}={\rm diag}(\rho_A, -p_\perp, -p_\perp, +p_\perp)\,,
\label{Tdiag}
\end{equation}
where
\begin{equation}
\rho_A\equiv\rho_{\rm kin}+V_A\;,\qquad
p_\perp\equiv\rho_{\rm kin}-V_A\;,
\label{rp}
\end{equation}
with
\begin{eqnarray}
\rho_{\rm kin} & \equiv & -\frac{1}{4}fF_{\mu\nu}F^{\mu\nu}
\;=\;\frac{1}{2}a^{-2}f\dot A^2,\label{rkin}\\
 & & \nonumber\\
V_A & \equiv & -\frac{1}{2}m^2A_\mu A^\mu
\;=\;\frac{1}{2}a^{-2}m^2A^2.\label{VA}
\end{eqnarray}
From Eq.~(\ref{Tdiag}) we see that the energy momentum tensor for our vector
field resembles the one of a perfect fluid, with the crucial difference that
the pressure along the longitudinal direction is of opposite sign to the 
pressure along the transverse directions. Thus, if the pressure is non-zero
and the vector field dominates the Universe, then large scale anisotropy will 
be generated. This is the reason we did not consider that $A_\mu$ can play the
role of the inflaton field in the first place.

However, in Ref.~\cite{vec} it was shown that, once \mbox{$m>H$}, the vector 
field undergoes quasi-harmonic oscillations, during which 
\mbox{$\overline{\rho_{\rm kin}}\approx\overline{V_A}$}, where the overline 
denotes average over a large number of oscillations.\footnote{Note that, since 
after inflation $\phi=M$ and $f=1$, the treatment and the results of 
Ref.~\cite{vec} are directly applicable here.} This result suggests that
\mbox{$\overline{p_\perp}\approx 0$} and the oscillating vector field behaves 
as isotropic pressureless matter. Therefore, it can indeed dominate the 
radiation background, {\em without introducing a large scale anisotropy}. 

In Ref.~\cite{vec} it was indeed confirmed that, during the 
oscillations, the density of the vector field scales as
\begin{equation}
\overline{\rho_A}\propto a^{-3}.
\label{rAosc}
\end{equation}
%
How does the density of the vector field scale before the onset of the 
oscillations? By virtue of Eqs.~(\ref{mlight}), (\ref{meff1}) and 
(\ref{massless}) we have that, 
\begin{equation}
\frac{m^2}{f}<H
\end{equation}
during inflation.
Then, it can be easily shown that Eq.~(\ref{EoMhom}) suggests that 
\mbox{$A\equiv|\mbox{\boldmath $A$}|$} remains frozen.
Hence, \mbox{$\rho_{\rm kin}\propto\dot A^2\rightarrow 0$}, while
\begin{equation}
\rho_A\simeq V_A\propto a^{-2},
\label{rfrz}
\end{equation}
where we considered Eq.~(\ref{VA}). As shown in Ref.~\cite{vec}, the scaling of
the vector density remains as such after inflation (when \mbox{$f=1$}) as
well, provided \mbox{$m<H(t)$}. Thus, we see that, despite the fact that
$A$ is frozen before the onset of the oscillations, the density of the vector 
field decreases.

\subsection{Curvaton Physics}

Using the results in the previous section we can trace the evolution of the
density of the vector field during and after inflation. As noted above, to 
avoid a large scale anisotropy, we need that the vector field 
begins oscillating before its decay and before it dominates the Universe.
Thus, we require:
\begin{equation}
\Gamma, m>\Gamma_A, H_{\rm dom}\;,
\label{Gs}
\end{equation}
where $\Gamma$ and $\Gamma_A$ are the decay rates of the inflaton field and
the vector curvaton field respectively and the subscript `dom' denotes the
time when the curvaton dominates the Universe (if it does not decay earlier).
Let us define the density parameter of the vector field as
\begin{equation}
\Omega\equiv\frac{\rho_A}{\rho}\,,
\label{omega}
\end{equation}
where $\rho$ is the background density typically corresponding to either the
oscillating inflaton field or the thermal bath of its decay products. 

In the standard picture, after the end of inflation the inflaton field 
undergoes quasi-harmonic oscillations until it decays at reheating. During 
these coherent oscillations the inflaton corresponds to a collection of
massive particles (inflatons) which behave like pressureless matter. Hence,
for the background density in this period we have \mbox{$\rho\propto a^{-3}$}.
After the decay of the inflaton (when \mbox{$\Gamma\geq H(t)$}) the Universe
becomes dominated by the relativistic decay products, in which case 
\mbox{$\rho\propto a^{-4}$}. In view of the above and Eqs.~(\ref{rAosc}) and
(\ref{rfrz}) it is easy to obtain the density parameter of the vector field
at the onset of its oscillations (denoted by `osc'):
\begin{equation}
\Omega_{\rm osc}\sim\Omega_{\rm end}\left(\frac{H_{\rm end}}{m}\right)^{2/3}
\min\left\{1,\frac{m}{\Gamma}\right\}^{-1/3}.
\label{Oosc}
\end{equation}
Similarly, if the curvaton decays before domination, we obtain 
\begin{equation}
\,\!
\Omega_{\rm dec}\sim\Omega_{\rm end}\left(\frac{H_{\rm end}}{m}\right)^{2/3}
\left(\frac{\Gamma}{\Gamma_A}\right)^{1/2}
\min\left\{1,\frac{m}{\Gamma}\right\}^{1/6},
\hspace{-.5cm}
\label{Odec}
\end{equation}
where `dec' denotes the time of the vector field decay 
(\mbox{$H_{\rm dec}=\Gamma_A$}). Finally, if the curvaton dominates the 
Universe before its decay (i.e. \mbox{$H_{\rm dom}>\Gamma_A$}) we find
\begin{equation}
H_{\rm dom}\sim\Omega_{\rm end}^2\Gamma\left(\frac{H_{\rm end}}{m}\right)^{4/3}
\min\left\{1,\frac{m}{\Gamma}\right\}^{1/3}.
\label{Hdom}
\end{equation}

Let us now estimate $\Omega_{\rm end}$. Since during inflation $A\simeq$~const.
we have \mbox{$\rho_A\simeq V_A=\frac{1}{2}m^2(A/a)^2$}. Hence, using that
\mbox{$\rho_{\rm end}=3H_{\rm end}^2m_P^2$}, we obtain
\begin{equation}
\Omega_{\rm end}
\sim e^{-2N_{\rm tot}}
\left(\frac{m}{H_{\rm end}}\right)^2\left(\frac{W_0}{m_P}\right)^2,
\label{Oend}
\end{equation}
where \mbox{$W_0\equiv A/a_0$} is the magnitude of the {\em physical} vector
field at the onset of inflation. 


In Ref.~\cite{vec} it is was explained that $A_\mu$ is the {\em comoving} 
vector field, which has the expansion of the Universe factored out. In a 
homogeneous and isotropic Universe the associated {\em physical} vector field 
is 
\begin{equation}
\mbox{\boldmath $W$}\equiv \mbox{\boldmath $A$}/a\,.
\label{W}
\end{equation}
This can be understood easily by considering the mass term in the Lagrangian in
Eq.~(\ref{L}). Using the flat FRW metric in Eq.~(\ref{FRW}) one has
\begin{equation}
\delta{\cal L}=\frac{1}{2}m^2A_\mu A^\mu=\frac{1}{2}m^2(A_t^2-a^{-2}A_iA_i)\,.
\end{equation}
Since the Lagrangian corresponds to a physical (observable) quantity we readily
see that the spatial components of the physical vector field are 
\mbox{$A_i/a$}, as in Eq.~(\ref{W}). Note also, that this is the explanation
of the explicit appearance of the scale factor in the results shown
in Eqs.~(\ref{PAflat}), (\ref{rkin}) and (\ref{VA}). For example, in view of 
Eq.~(\ref{PA}), the value of the scale invariant power spectrum 
of the physical vector field $W_\mu$ is 
\mbox{${\cal P_W}={\cal P_A}/a^2=(H/2\pi)^2$}, i.e. identical to the case of a 
massless scalar field \cite{vec}. 

From the above we see that, even though $A$ is frozen during inflation, 
\mbox{$W\equiv|\mbox{\boldmath $W$}|=A/a$} is gradually decreasing, which 
explains the exponential suppression of $\Omega_{\rm end}$ in Eq.~(\ref{Oend}).

\subsection{The curvature perturbation}

The curvature perturbation associated with the vector field is
\begin{equation}
\zeta_A=-H\frac{\delta\rho_A}{\dot\rho_A}=
\frac{1}{3}\left.\frac{\delta\rho_A}{\rho_A}\right|_{\rm dec},
\label{zA}
\end{equation}
where we considered that, before its decay, the vector field is undergoing
coherent oscillations, for which \mbox{$\dot\rho_A=-3H\rho_A$} as suggested by
Eq.~(\ref{rAosc}). Since during oscillations we have 
\mbox{$\overline{\rho_A}\approx2\overline{V_A}=a^{-2}m^2\overline{A^2}$}, 
we find
\begin{equation}
\zeta_A=\left.\frac{\delta\rho_A}{3\rho_A}\right|_{\rm dec}\simeq
\frac{2}{3}\left.\frac{\delta\hat A}{\hat A}\right|_{\rm dec}
\approx\frac{2}{3}\left.\frac{\delta A}{A}\right|_{\rm osc}\;,
\label{zAosc}
\end{equation}
where we took into account that, during the oscillations, both $\delta A$
and $A$ obey the same equation of motion, since Eq.~(\ref{EoMhom}) is linear.
We also considered that 
\mbox{$\overline{A^2}\approx\frac{1}{2}\hat A^2$}, where by $\hat A$ we denote 
the amplitude of the oscillations, which is equal to $A$ at the onset of the 
oscillations.

Before the onset of the oscillations we have \mbox{$m/f<H$}, which means that
$A$ is frozen. However, as evident from Eq.~(\ref{PAflat}), $\delta A$ grows
as \mbox{$\delta A\propto a$}. That is, {\em although the spectrum of the 
perturbations of the vector field is scale invariant, its amplitude grows with 
the scale factor of the Universe}. This implies that
\begin{equation}
\left.\frac{\delta A}{A}\right|_{\rm osc}\!\!=
\frac{a_{\rm osc}}{a_*}
\left.\frac{\delta A}{A}\right|_*=
\left(\frac{a_{\rm osc}}{a_*}\right)
\frac{a_*H_*}{2\pi A_*}\approx\frac{H_*}{2\pi W_{\rm osc}}\,,
\label{dA/A}
\end{equation}
where we have used that 
\mbox{$W_{\rm osc}\equiv(A/a)_{\rm osc}\approx A_*/a_{\rm osc}$} and
we have assumed that \mbox{$\delta A/A<1$} at all times.
The above shows that the growth of the amplitude of the perturbations
before the onset of the oscillations is due to the decrease of the physical
vector field, according to Eq.~(\ref{W}). Using Eq.~(\ref{W}), it is easy to
find
\begin{equation}
W_{\rm osc}\sim W_0\,e^{-N_{\rm tot}}\left(\frac{m}{H_{\rm end}}\right)^{2/3}
\min\left\{1,\frac{m}{\Gamma}\right\}^{-1/6},
\label{Wosc}
\end{equation}
where we assumed that $A$ is frozen throughout inflation. Putting together
Eqs.~(\ref{zAosc}), (\ref{dA/A}) and (\ref{Wosc}) we obtain
\begin{equation}
\zeta_A
\sim e^{N_{\rm tot}}\frac{H_*}{W_0}
\left(\frac{H_{\rm end}}{m}\right)^{2/3}
\min\left\{1,\frac{m}{\Gamma}\right\}^{1/6}.
\label{z}
\end{equation}

\subsection{The parameter space}\label{ps}

Substituting from the above $e^{N_{\rm tot}}$ into Eq.~(\ref{Oend}) we get
\begin{equation}
\Omega_{\rm end}
\sim\zeta_A^{-2}\left(\frac{H_*}{m_P}\right)^2
\left(\frac{m}{H_{\rm end}}\right)^{2/3}
\min\left\{1,\frac{m}{\Gamma}\right\}^{1/3},
\label{Oend1}
\end{equation}
which, remarkably, is independent of $W_0$. Plugging Eq.~(\ref{Oend1}) into
Eqs.~(\ref{Odec}) and (\ref{Hdom}) we find that, 
if the vector curvaton decays before domination
\begin{equation}
\Omega_{\rm dec}
\sim\zeta_A^{-2}\left(\frac{H_*}{m_P}\right)^2
\left(\frac{\Gamma}{\Gamma_A}\right)^{1/2}
\min\left\{1,\frac{m}{\Gamma}\right\}^{1/2},
\label{Odec1}
\end{equation}
while if the vector curvaton dominates the 
Universe before its decay
\begin{equation}
H_{\rm dom}
\sim\Gamma\,\zeta_A^{-4}
\left(\frac{H_*}{m_P}\right)^4
\min\left\{1,\frac{m}{\Gamma}\right\}\,.
\label{Hdom1}
\end{equation}

Solving Eqs.~(\ref{Odec1}) and (\ref{Hdom1}) for $H_*$ we obtain
\begin{equation}
\frac{H_*}{m_P}\sim\frac{\zeta}{\sqrt{\Omega_{\rm dec}}}
\left(\frac{\max\{H_{\rm dom},\Gamma_A\}}{\min\{\Gamma, m\}}\right)^{1/4},
\label{H*}
\end{equation}
where we used the fact that, in the curvaton mechanism 
\mbox{$\zeta\sim\Omega_{\rm dec}\zeta_A$},
where \mbox{$\zeta\simeq 5\times 10^{-5}$} is the observed curvature 
perturbation. Now, considering that
\mbox{$\Omega_{\rm dec}\lsim 1$}, 
\mbox{$\max\{H_{\rm dom},\Gamma_A\}\geq\Gamma_A$} and 
\mbox{$m\leq 0.1\,H_*$}, it can be easily verified that
\begin{equation}
\left(\frac{H_*}{m_P}\right)^5\geq
10\,\zeta^4\,\frac{\Gamma_A}{m_P}\,.
\label{Hbound}
\end{equation}
The lower bound in the above is attained when 
\mbox{$\Gamma\geq m$}, \mbox{$m\rightarrow 0.1\,H_*$}, 
\mbox{$\Omega_{\rm end}\rightarrow 1$}  and 
\mbox{$H_{\rm dom}\rightarrow\Gamma_A$}.
This case corresponds to almost prompt reheating and curvaton decay as soon as 
the latter dominates the Universe. 

Demanding that the decay of the curvaton occurs before Big Bang Nucleosynthesis
(BBN) imposes the bound \mbox{$\Gamma_A>T_{\rm BBN}^2/m_P$}, which suggests
\begin{equation}
\frac{H_*}{m_P}>
10^{1/5}
\zeta^{4/5}\left(\frac{T_{\rm BBN}}{m_P}\right)^{2/5}\Rightarrow
H_*>10^6\,{\rm GeV}\,,
\label{H*bound}
\end{equation}
where \mbox{$T_{\rm BBN}\simeq 1\;$MeV} is the temperature at BBN.
Hence, under this mechanism, the inflationary energy scale cannot be lower than
\mbox{$V_*^{1/4}\sim 10^{12}\,$GeV}, which agrees with the generic bound for
the curvaton mechanism \cite{Hbound}. 

However, there is an important subtlety that needs to be considered here.
Even though $W_0$ drops our from the calculations, one still must take into
account the evolution of $W=A/a$ during inflation. This is because, in the
above, we have assumed that $W$ decreases as $W\propto a^{-1}$ since $A$ is 
frozen. However, because \mbox{${\cal P_W}\simeq(H_*/2\pi)^2$}, the decrease
of $W$ will be halted if \mbox{$W\lsim H_*$}. Thus, we need to postulate that
\mbox{$W_{\rm end}>H_{\rm end}\approx H_*$}.\footnote{\label{foot}
Note that this bound can be much more stringent if $A$ is not frozen but
diminishes with time. This is why we have chosen \mbox{$\alpha=3$} in 
Sec.~\ref{hvf}.} 
For this we need to obtain an estimate of $N_{\rm tot}$.

As discussed above, the parameter space for $H_*$ is maximised if the vector 
curvaton decays when it is about to dominate the density of the Universe. This 
means that, after the decay of the inflaton field, the Universe remains 
radiation dominated, in which case, the number of e-folds corresponding to the
horizon at present is
\begin{equation}
N_H\simeq 67-
\frac{1}{2}\ln\left(\frac{m_P}{H_*}\right)+
\ln\left(\frac{H_*}{\Gamma}\right).
\label{NH}
\end{equation}
The parameter space for $H_*$ is maximised when \mbox{$\Gamma\sim H_*$},
which also results in minimising $N_H$. Now, postulating
\mbox{$W_{\rm end}=e^{-N_{\rm tot}}W_0>H_*$} and considering 
\mbox{$N_{\rm tot}>N_H$} (so that inflation solves the horizon and flatness 
problems) we obtain the bound: \mbox{$W_0>10^8\,m_P$}! Such huge values of
$W_0$ are unacceptably unrealistic. If, on the other hand, we demand 
\mbox{$W_0\lsim m_P$} then we find that the bound 
\mbox{$W_{\rm end}>H_{\rm end}$} requires
\begin{equation}
N_{\rm tot}\leq\ln\left(\frac{W_0}{H_*}\right)
\lsim\ln\left(\frac{m_P}{H_*}\right).
\label{Ntotbound}
\end{equation}
In view of Eqs.~(\ref{H*bound}) and (\ref{NH}), the above bound cannot satisfy
\mbox{$N_{\rm tot}>N_H$}, i.e. inflation is not enough to solve the horizon 
and flatness problems. It seems, therefore, that 
some modification is required, which will allow low scale inflation,
for the vector curvaton scenario to work.

\section{Mass increment}

In Ref.~\cite{low} the possibility of low scale inflation in the context of the
curvaton mechanism was investigated. It was shown that this is indeed possible
in two ways. One possibility is to consider as curvaton a pseudo-Nambu 
Goldstone boson, whose order parameter increases after the cosmological scales 
exit the horizon during inflation. This mechanism was implemented in 
Ref.~\cite{laza} and it was shown that inflation with $H_*$ at least as low as 
1~TeV was possible to attain. The other technique involves a phase transition 
after the end of inflation, which gives rise to a sudden increment of the 
curvaton's mass (see also Ref.~\cite{low1}).
It is this mechanism that we attempt to implement in this paper to the case of
a vector curvaton.

We assume, therefore, that a phase transition takes place at some time after
the end of inflation but before the onset of the vector field oscillations.
The mass of the vector field is increased from $m$ to $m_0$ at this phase 
transition to become larger than the Hubble scale at the time, so that 
oscillations
begin immediately. Hence, the phase transition corresponds to Hubble scale
\mbox{$m<H_{\rm osc}\leq m_0$}.

\subsection{Relaxing the bound on the inflationary scale}

The sudden increment of the mass of the vector field results in a corresponding
growth of the density of the vector field. Since, 
\mbox{$\rho_A\simeq V_A\propto m^2$} before the oscillations (c.f. 
Eq.~(\ref{VA})), we find that $\Omega_{\rm osc}$ grows by a factor of 
$(m_0/m)^2$. Hence, we have
\begin{equation}
\,\!
\Omega_{\rm osc}\sim\Omega_{\rm end}
\left(\frac{m_0}{m}\right)^2
\left(\frac{H_{\rm end}}{H_{\rm osc}}\right)^{2/3}
\min\left\{1,\frac{H_{\rm osc}}{\Gamma}\right\}^{-1/3}
\!\!
.
\hspace{-.5cm}
\label{Oosc+}
\end{equation}
The above is directly obtainable by Eq.~(\ref{Oosc}) with the substitution 
\mbox{$m\rightarrow H_{\rm osc}$} and taking also the growth factor $(m_0/m)^2$
into account. An important constraint here is that the increment of the density
of the vector field does not surpass the overall density available at the phase
transition, i.e.
\begin{equation}
\Omega_{\rm osc}\leq 1.
\label{osccons}
\end{equation}

Using the above, in the case when 
the curvaton decays before domination, we obtain 
\begin{eqnarray}
\Omega_{\rm dec} & \sim & \Omega_{\rm end}
\left(\frac{m_0}{m}\right)^2
\left(\frac{H_{\rm end}}{H_{\rm osc}}\right)^{2/3}
\left(\frac{\Gamma}{\Gamma_A}\right)^{1/2}\times
\nonumber\\
& & \times\min\left\{1,\frac{H_{\rm osc}}{\Gamma}\right\}^{1/6},
\label{Odec+}
\end{eqnarray}
while in the case when the curvaton dominates the 
Universe before its decay
\begin{equation}
\,\!
H_{\rm dom}\sim\Omega_{\rm end}^2
\Gamma
\left(\frac{m_0}{m}\right)^4
\left(\frac{H_{\rm end}}{H_{\rm osc}}\right)^{4/3}
\!\!
\min\left\{1,\frac{H_{\rm osc}}{\Gamma}\right\}^{1/3}
\!\!\!
.
\hspace{-.5cm}
\label{Hdom+}
\end{equation}
Note that, in the above, 
$\Omega_{\rm end}$ is still given by Eq.~(\ref{Oend}).

It is easy to see that Eqs.~(\ref{Wosc}) and (\ref{z}) remain unaffected
by the mass increment, apart from the substitution 
\mbox{$m\rightarrow H_{\rm osc}$}. Thus, we have
\begin{equation}
\!\,W_{\rm osc}\sim W_0\,e^{-N_{\rm tot}}
\left(\frac{H_{\rm osc}}{H_{\rm end}}\right)^{2/3}
\min\left\{1,\frac{H_{\rm osc}}{\Gamma}\right\}^{-1/6},
\hspace{-1cm}
\label{Wosc+}
\end{equation}
and
\begin{equation}
\zeta_A
\sim e^{N_{\rm tot}}\frac{H_*}{W_0}
\left(\frac{H_{\rm end}}{H_{\rm osc}}\right)^{2/3}
\min\left\{1,\frac{H_{\rm osc}}{\Gamma}\right\}^{1/6}.
\label{z+}
\end{equation}

Then, working as in the previous section, we obtain
\begin{eqnarray}
\Omega_{\rm end} & \sim & \zeta_A^{-2}\left(\frac{H_*}{m_P}\right)^2
\left(\frac{H_{\rm osc}}{H_{\rm end}}\right)^{2/3}
\left(\frac{m}{H_{\rm osc}}\right)^2\times
\nonumber\\
 & & \times\min\left\{1,\frac{H_{\rm osc}}{\Gamma}\right\}^{1/3},
\label{Oend1+}
\end{eqnarray}
which is the equivalent of Eq.~(\ref{Oend1}). Using this we find
that, if the vector curvaton decays before domination
\begin{eqnarray}
\Omega_{\rm dec} & \sim & \zeta_A^{-2}\left(\frac{m_0}{H_{\rm osc}}\right)^2
\left(\frac{H_*}{m_P}\right)^2
\left(\frac{\Gamma}{\Gamma_A}\right)^{1/2}\times
\nonumber\\
 & & \times\min\left\{1,\frac{H_{\rm osc}}{\Gamma}\right\}^{1/2},
\label{Odec1+}
\end{eqnarray}
while if the vector curvaton dominates the 
Universe before its decay
\begin{equation}
H_{\rm dom}
\sim\Gamma\,\zeta_A^{-4}\left(\frac{m_0}{H_{\rm osc}}\right)^4
\left(\frac{H_*}{m_P}\right)^4
\min\left\{1,\frac{H_{\rm osc}}{\Gamma}\right\}\,.
\label{Hdom1+}
\end{equation}

Solving Eqs.~(\ref{Odec1+}) and (\ref{Hdom1+}) for $H_*$ we obtain
\begin{equation}
\frac{H_*}{m_P}\sim\frac{\zeta}{\sqrt{\Omega_{\rm dec}}}
\left(\frac{\max\{H_{\rm dom},\Gamma_A\}}%
{\min\{\Gamma,H_{\rm osc}\}}\right)^{1/4}
\frac{H_{\rm osc}}{m_0},
\label{H*+}
\end{equation}
where we used the fact that, in the curvaton mechanism 
\mbox{$\zeta\sim\Omega_{\rm dec}\zeta_A$}. Comparing the above with 
Eq.~(\ref{H*}) we see that, apart from the substitution 
\mbox{$m\rightarrow H_{\rm osc}$}, there is an extra factor of
$H_{\rm osc}/m_0$ in the right-hand-side. This means that, if 
\mbox{$m_0\gg H_{\rm osc}$}, the lower bound on $H_*$ can be substantially 
relaxed to the desired level. 

In contrast to the previous section, due to the extra factor of 
$H_{\rm osc}/m_0$, the lower bound on $H_*$ can be more relaxed the 
{\em latter} the oscillations begin. Hence, the lowest bound is found when 
\mbox{$H_{\rm osc}\rightarrow\Gamma_A$}. Indeed, in this case it is easy to 
find
\begin{equation}
\frac{H_*}{m_P}>\zeta\,\frac{\Gamma_A}{m_0}\,.
\label{Hbound+}
\end{equation}
The above shown bound corresponds to \mbox{$\Omega_{\rm end}\rightarrow 1$}  
and \mbox{$\Gamma\geq H_{\rm osc}>\Gamma_A\geq H_{\rm dom}$}, i.e. to the case 
when the phase transition (which results in the oscillations of the
vector curvaton) takes place just before the latter decays and as soon as it 
dominates the Universe. Since we need some oscillations before the curvaton 
domination and decay, in order to avoid a long-range anisotropy, the above 
lower bound is unattainable. 

\subsection{Additional bound on the inflationary scale}

The decay rate of the vector curvaton is
\begin{equation}
\Gamma_A\sim h^2m_0\quad{\rm with}\quad
\frac{m_0}{m_P}\lsim h\lsim 1\,
\label{GA}
\end{equation}
where the lower bound to the decay coupling $h$ corresponds to gravitational 
decay.
%
Using this 
we have
\begin{equation}
\frac{\max\{H_{\rm dom},\Gamma_A\}}{H_{\rm osc}}\geq
\frac{\Gamma_A}{H_{\rm osc}}\gsim
\left(\frac{m_0}{m_P}\right)^2\frac{m_0}{H_{\rm osc}}\,.
\end{equation}
Inserting the above into Eq.~(\ref{H*+}) and after a little algebra we obtain
\begin{equation}
\frac{H_*}{m_P}\gsim\frac{\zeta^2}{\Omega_{\rm dec}}
\frac{H_{\rm osc}}{H_*}\sqrt{\frac{H_{\rm osc}}{m_0}}\;
\max\left\{1,\frac{H_{\rm osc}}{\Gamma}\right\}^{1/2},
\label{H*bound+1}
\end{equation}
where the lower bound is attained when the vector curvaton decays 
gravitationally.

Now, from Eqs.~(\ref{osccons}) and (\ref{Oend1+}) we get
\begin{equation}
\frac{m_0}{m_P}\lsim\frac{\zeta}{\Omega_{\rm dec}}\frac{H_{\rm osc}}{H_*}\,,
\label{m0bound}
\end{equation}
where we used \mbox{$\zeta_A\sim\Omega_{\rm dec}\zeta$}. The upper bound 
corresponds the the case when the density of the oscillating vector field 
dominates the Universe immediately after the phase transition. Using the above,
Eq.~(\ref{H*bound+1}) results in the bound
\begin{equation}
\frac{H_*}{m_P}\gsim\frac{\zeta^3}{\Omega_{\rm dec}}
\left(\frac{H_{\rm osc}}{H_*}\right)^2
\max\left\{1,\frac{H_{\rm osc}}{\Gamma}\right\}.
\label{H*bound+}
\end{equation}

The above bound suggests that the mass increment mechanism can relax the lower 
bound on $H_*$ {\em only if the phase transition occurs much later than the 
end of inflation}. To show this, consider the opposite case, when 
\mbox{$H_{\rm osc}\sim H_*\geq\Gamma$}. In this case, and considering also
that \mbox{$\Omega_{\rm dec}\leq 1$}, we find
\begin{equation}
H_*\gsim\zeta^3m_P\sim 10^5\,{\rm GeV}\,,
\end{equation}
which is not too different from the bound in Eq.~(\ref{H*bound}).

\subsection{The parameter space revisited}

Let us investigate now whether, under the mass increment mechanism, it
is possible to achieve enough e-folds of inflation to solve the horizon and 
flatness problems while generating the observed amplitude for the curvature
perturbation. To maximise the parameter space, we assume 
\mbox{$\Omega_{\rm dec}\rightarrow 1$}, i.e. \mbox{$\zeta_A\rightarrow\zeta$}.
Also, we consider \mbox{$\Gamma_A\geq H_{\rm dom}$}, which means that the
vector curvaton decays as soon as it dominates the Universe. Finally, since
the bounds on the inflationary scale are relaxed for small values of 
$H_{\rm osc}$, we assume \mbox{$\Gamma\geq H_{\rm osc}$}, that is the phase 
transition occurs after the decay of the inflaton field.

Under the above assumptions Eq.~(\ref{z+}) gives
\begin{equation}
e^{N_{\rm tot}}\sim\zeta\,\frac{W_0}{H_*}
\left(\frac{H_{\rm osc}}{H_*}\right)^{2/3}
\left(\frac{\Gamma}{H_{\rm osc}}\right)^{1/6},
\label{Ntot1}
\end{equation}
while the bound in Eq.~(\ref{H*bound+}) can be written as
\begin{equation}
\frac{H_*}{m_P}\gsim\zeta
\left(\frac{H_{\rm osc}}{m_P}\right)^{2/3}.
\label{H*bound+2}
\end{equation}
Now, writing Eq.~(\ref{NH}) as
\begin{equation}
e^{N_H}\sim 10^{29}
\left(\frac{H_*}{m_P}\right)^{1/2}
\frac{H_*}{\Gamma}
\label{NH1}
\end{equation}
and using Eqs.~(\ref{Ntot1}) and (\ref{H*bound+2}), the requirement 
\mbox{$N_{\rm tot}\geq N_H$} results in the bound:
\begin{equation}
H_{\rm osc}\lsim 10^{-30}
\left(\frac{W_0}{m_P}\right)^{6/5}
\left(\frac{\Gamma}{H_*}\right)^{7/5}m_P\;.
\label{Hoscbound}
\end{equation}
Taking \mbox{$W_0\sim m_P$} and also \mbox{$\Gamma\sim H_*$} (prompt reheating)
we find that the phase transition, which results to the growth of the mass of
the vector field, can take place at temperature
\begin{equation}
T_{\rm osc}\lsim 1\;{\rm TeV}\,.
\end{equation}
The above upper bound can be saturated when the bound in Eq.~(\ref{H*bound+2})
is saturated, i.e. when the vector curvaton decays gravitationally. 

Thus, we 
see that it is indeed possible to attain enough inflation to solve the horizon 
and flatness problems and explain the curvature perturbations in the Universe,
when the phase transition, which results in mass increment for the vector 
field, occurs around the time of the breakdown of electroweak unification.

\section{Scalar Field Concerns}

Apart from the above considerations there are a couple of issues regarding
the scalar field $\phi$, whose evolution is crucial during inflation, since
it controls $f(\phi)$.

\subsection{\boldmath Production of $\phi$ during inflation}

One issue that needs to be examined is whether $\phi$ also manages to
obtain a superhorizon spectrum of perturbations and, if so, whether they
may give rise to an acceptable or not contribution to the curvature 
perturbation.

Being tachyonic, $\phi$ is guaranteed to undergo particle production during 
inflation. One can understand this as follows. From Eq.~(\ref{EoMphi}) one
obtains the following equation of motion for the Fourier modes of the
perturbation $\delta\phi$ of the field 
\begin{eqnarray}
 & \left[\partial_t^2+3H\partial_t-m_\phi^2+\left(\frac{k}{a}\right)^2\right]
\delta\varphi_k=0\,,
 & \label{eomphi}
\end{eqnarray}
where 
\begin{equation}
\delta\phi(t, \mbox{\boldmath $x$})=
\int\frac{d^3k}{(2\pi)^{3/2}}\;\delta\varphi_k 
(t, \mbox{\boldmath $k$})
\,
e^{i\mbox{\scriptsize\boldmath $k\cdot x$}}.
\label{fourier1}
\end{equation}
Solving Eq.~(\ref{eomphi}) with vacuum boundary conditions in the same manner 
as in Sec.~\ref{pp} one obtains the following power spectrum
\begin{equation}
{\cal P}_\phi\approx\frac{4\pi}{|\Gamma(1-\nu)|^2}
\left(\frac{H}{2\pi}\right)^2\left(\frac{k}{2aH}\right)^{3-2\nu}
\label{Pphi}
\end{equation}
where
\begin{equation}
\nu\equiv
\sqrt{\frac{9}{4}+\left(\frac{m_\phi}{H}\right)^2}
=\frac{3}{2}\left(\frac{1}{n}+1\right),
\label{vphi}
\end{equation}
where we used Eq.~(\ref{mphiH}), taking \mbox{$\alpha=3$} as discussed in
Sec.~\ref{sp}. From Eq.~(\ref{Pphi}) it is evident that a scale invariant
spectrum is attainable only if \mbox{$\nu\approx 3/2$}. However, as suggested 
by Eq.~(\ref{vphi}), such a spectrum is attainable only if $n$ is very large.
For example, if \mbox{$n=1$}, as discussed in Sec.~\ref{hvf}, then 
\mbox{$\nu=5/2$} and \mbox{${\cal P}_\phi\propto k^{-2}$}. If such a spectrum
of perturbations contributed significantly to the curvature perturbation
then it would be incompatible with the observations.

The contribution of the perturbations of $\phi$ to the curvature perturbation 
is
\begin{equation}
\delta\zeta\sim\Omega_\phi\zeta_\phi\lsim\Omega_\phi\;,
\label{dz}
\end{equation}
where we considered that \mbox{$\zeta_\phi\lsim 1$} and also defined the
density parameter of $\phi$ as
\begin{equation}
\Omega_\phi\equiv\frac{\rho_\phi}{\rho}\sim\left(\frac{M}{m_P}\right)^2,
\label{Ophi}
\end{equation}
where we used that, during inflation 
\mbox{$\rho_\phi\sim V_0\sim(m_\phi M)^2$} [c.f. Eq.~(\ref{V0})] and 
\mbox{$(m_\phi/H)^2=\frac{9}{4n}(\frac{1}{n}+2)\sim{\cal O}(1)$},
according to Eq.~(\ref{mphiH}). 

We need to make sure that $\phi$ does not produce an excessive curvature
perturbation compared to the observations, which suggest 
\mbox{$\zeta\simeq 5\times 10^{-5}$}. Thus, avoiding conflict with observations
is guaranteed if \mbox{$\delta\zeta\lsim\zeta$}, i.e.
\begin{equation}
M\lsim \sqrt\zeta\,m_P\;\sim 2\times 10^{16}\,{\rm GeV}\,.
\label{Mbound}
\end{equation}

Note that, as the roll of $\phi$ towards $M$ progresses, the associated 
curvature perturbation \mbox{$\zeta_\phi\propto\delta\phi/\phi$} is diminished,
not only because $\phi$ grows but also because the spectrum of $\delta\phi$ is
red. Hence, the above bound on $M$ can be relaxed if the cosmological scales
exit the horizon after the initial outburst of tachyonic perturbations has
subsided somewhat.\footnote{\label{MmP}
Note that, if \mbox{$M\ll m_P$} then 
$\phi$ cannot play the role of the inflaton because \mbox{$V_0\ll V_*$}, where
we used Eq.~(\ref{V0}) and also that \mbox{$m_\phi\sim H_*$}, according to
Eq.~(\ref{mphiH}).}

\subsection{\boldmath Source terms in the field equation of $\phi$}

The dependence on $\phi$ of the vector field kinetic term gives rise to
source terms in the field equation of the scalar field. To study their 
influence let us consider the following Lagrangian density:
\begin{eqnarray}
{\cal L} & = & -\frac{1}{4}f(\phi)F_{\rm \mu\nu}F^{\mu\nu}+
\frac{1}{2}m^2A_\mu A^\mu+\nonumber\\
 & & +\frac{1}{2}D_\mu\phi(D^\mu\phi)^*-V(\phi)\,,
\label{L2}
\end{eqnarray}
where $V(\phi)$ is given by Eq.~(\ref{Vphi}) and
\mbox{$D_\mu\phi=\partial_\mu\phi-igA_\mu\phi$}. The case of constant mass
corresponds to \mbox{$g=0$}, while the Higgsed vector field case
corresponds to \mbox{$m=0$}. From the above we find
\begin{eqnarray}
& & 
\hspace{-1.4cm}
\ddot\phi+3H\dot\phi-m_\phi^2\phi=
-\frac{1}{4}f'(\phi)
F_{\mu\nu}F^{\mu\nu}+g^2A_\mu A^\mu
\;\Rightarrow
\nonumber\\
& & 
\hspace{-1.4cm}
\ddot\phi+3H\dot\phi=\nonumber\\
& & 
\hspace{-1.4cm}
\left\{m_\phi^2-a^{-2}\left[(gA)^2-n\left(\frac{\phi}{M}\right)^{2(n-1)}
\left(\frac{\dot A}{M}\right)^2\right]\right\}\phi\,,
\label{EoMphiA}
\end{eqnarray}
where the prime denotes derivative with respect to $\phi$ and, in the last line
of the above we have used Eq.~(\ref{fn}) as well as that 
\mbox{$A_\mu A^\mu=-a^{-2}A^2$} and also 
\mbox{$F_{\mu\nu}F^{\mu\nu}=-2a^{-2}\dot A^2$}.

Assuming \mbox{$\alpha=3$} as discussed in Sec.~\ref{sp}, Eq.~(\ref{mphiH})
suggests \mbox{$m_\phi=\frac{\sqrt{1+3n}}{n}H\sim H$}. On the other hand,
due to Eqs.~(\ref{mlight}), (\ref{meff1}) and (\ref{massless}), we have 
\mbox{$m^2/f<H$}, which means that \mbox{$A\equiv|\mbox{\boldmath $A$}|$}
is frozen, since the mass term in Eq.~(\ref{EoMhom}) is negligible compared
to the ``friction'' term $3H\dot A$. This implies that 
\mbox{$\dot A\rightarrow 0$}, which means that Eq.~(\ref{EoMphiA}) can
be recast as
\begin{equation}
\ddot\phi+3H\dot\phi+(g^2 W^2-m_\phi^2)\phi\simeq 0\,,
\label{EoMphiB}
\end{equation}
where we used also Eq.~(\ref{W}). Now, in the constant mass case 
\mbox{$g=0$} and, therefore, the above equation reduces to
Eq.~(\ref{EoMphi}). In the Higgsed vector field case, though, this is not 
necessarily so. Indeed,
due to Eq.~(\ref{massless}), we have \mbox{$gW\lsim 0.1 H(W/M)$}, which
might still dominate \mbox{$m_\phi\sim H$}, if \mbox{$W>{\cal O}(10)M$}.
Since, in principle, \mbox{$W\leq W_0\lsim m_P$} this is not impossible, given
Eq.~(\ref{Mbound}).

What happens if \mbox{$gW>m_\phi$}? Note, at first, that, since $A$ is frozen
during inflation, \mbox{$W\propto a^{-1}$}, i.e. $W$ is decreasing 
exponentially, which means that eventually $m_\phi$ becomes dominant, whatever 
the initial value of $W$. Still, just after the onset of inflation we may well 
have \mbox{$gW_0\gg m_\phi\sim H$}. According to Eq.~(\ref{EoMphiB}), a 
positive mass-squared larger than $H$ would rapidly send $\phi$ to the origin. 
Consequently, since \mbox{$m=0$} in Eq.~(\ref{L2}) in the Higgsed vector 
field case, the vector field is rendered exactly massless. This means that
conformal invariance is restored and no perturbations of the vector field
are generated \cite{vec}. 

As $W$ decreases, however, the effective mass-squared of $\phi$:
\mbox{$m_{\rm eff}^2\equiv g^2W^2-m_\phi^2$} becomes smaller than $H^2$, in 
which case particle production of $\phi$ generates a condensate for $\phi$ of 
order \mbox{$\langle\phi^2\rangle\sim (H/2\pi)^2$}. Indeed, Eq.~(\ref{vphi}) 
becomes \mbox{$\nu=\sqrt{\frac{9}{4}+(m_\phi/H)^2-(gW/H)^2}$}. Hence, particle 
production begins when \mbox{$(gW/H)^2<\frac{9}{4}+\frac{1+3n}{n^2}$}, where 
we used Eq.~(\ref{mphiH}) with \mbox{$\alpha=3$}. After 
\mbox{$m_{\rm eff}^2<0$}, a phase transition sends $\phi$ rolling off the 
origin and down the potential hill in Eq.~(\ref{Vphi}) as described in 
Sec.~\ref{frsf}, while Eq.~(\ref{EoMphiB}) reduces to Eq.~(\ref{EoMphi}).

Thus, when \mbox{$gW_0>H_*$}, there is an initial period of inflation, where
there is no vector particle production, while $\phi$ is sent to the origin.
This period lasts for
\begin{equation}
N_W=\ln\left(\frac{gW_0}{H_*}\right) 
\label{NW}
\end{equation}
e-foldings. Afterwards, a phase transition occurs which releases $\phi$ 
from the origin, the conformal invariance of the vector field is broken and
particle production takes place as discussed in Sec.~\ref{hvf}. From the above
we see that the Higgs vector field case has the advantage of explaining the
initial condition of $\phi$ on top of the potential hill, if $W_0$ is large 
enough.

\section{A concrete example}

To visualise the above findings we briefly study a particular example,
considering the case of a Higgsed vector field. Thus, the Lagrangian density 
is given in Eq.~(\ref{L1}). We take \mbox{$\alpha=3$} and \mbox{$n=1$}, that is
we assume that $f(\phi)$ is given by Eq.~(\ref{f2}). 
According to Eq.~(\ref{phia}) \mbox{$\phi\propto a$}, while Eq.~(\ref{Nphi}) 
suggests
\begin{equation}
N_\phi=\ln\left(\frac{M}{\phi_0}\right).
\label{Nphi1}
\end{equation}
In order to obtain an approximately scale invariant spectrum of perturbations
we have to take the constraint in Eq.~(\ref{massless}) into account. 
We assume that this constraint well satisfied, so that
\begin{equation}
gM\ll 0.1\,H\,.
\label{gMH}
\end{equation}

For the scalar field, which controls the mass of the vector field 
\mbox{$m(\phi)=g\phi$}, we consider a Higgs-type potential
\begin{equation}
V(\phi)=\frac{1}{4}\lambda(\phi^2-M^2)^2,
\label{higgs}
\end{equation}
with $\lambda$ being a constant. In view of the above potential and also 
of Eq.~(\ref{mphiH}), we have
\begin{equation}
m_\phi=\sqrt\lambda M=2H_*\;.
\label{mH}
\end{equation}
From Eqs.~(\ref{gMH}) and (\ref{mH}) we readily obtain
\begin{equation}
g\ll\sqrt\lambda/20\,.
\label{gl}
\end{equation}

We assume that reheating is prompt and also that the vector curvaton decays as 
soon as it dominates the Universe. This means
\begin{equation}
\Gamma\sim H_*
\quad{\rm and}\quad
\Omega_{\rm dec}\sim 1
\quad{\rm and}\quad
\Gamma_A\geq H_{\rm dom}\;.
\label{assum}
\end{equation}
Also, we assume that the vector curvaton decays through gravitational 
interactions, i.e.
\begin{equation}
\Gamma_A\sim\frac{m_0^3}{m_P^2}\,.
\label{GAex}
\end{equation}
Furthermore, we assume that \mbox{$W_0\sim m_P$}. This means that the bound in
Eq.~(\ref{Hoscbound}) becomes \mbox{$H_{\rm osc}\lsim 10^{-30}m_P$}. 

We choose the following value for our example
\begin{equation}
H_{\rm osc}\sim 10^{-32}m_P\;\Rightarrow\;T_{\rm osc}\sim 100\,{\rm GeV}\,,
\label{Hoscex}
\end{equation}
i.e. the phase transition which results in the increment of the mass of the
vector filed occurs at the breaking of electroweak unification.

Similarly, we choose the decay rate of the vector curvaton to be
\begin{equation}
\Gamma_A\sim 10^{-36}m_P\;\Rightarrow\;T_{\rm dec}\sim 1\,{\rm GeV}\gg 
T_{\rm BBN}\;.
\label{GAex1}
\end{equation}
From Eqs.~(\ref{GAex}) and (\ref{GAex1}) we find
\begin{equation}
m_0\sim 10^{-12}m_P\sim 10^6\,{\rm GeV}\,.
\label{m0ex}
\end{equation}

Now, in view of Eq.~(\ref{assum}), Eqs.~(\ref{Oosc+})  and (\ref{Oend1+}) give
\begin{equation}
\Omega_{\rm osc}\sim\zeta^{-2}
\left(\frac{H_*}{m_P}\right)^2
\left(\frac{m_0}{H_{\rm osc}}\right)^2.
\end{equation}
Similarly, Eq.~(\ref{H*+}) becomes
\begin{equation}
\frac{H_*}{m_P}\sim\zeta\left(\frac{\Gamma_A}{H_{\rm osc}}\right)^{1/4}
\frac{H_{\rm osc}}{m_0}\,.
\label{H*consex}
\end{equation}
Combining the above we find
\begin{equation}
\Omega_{\rm osc}\sim
\left(\frac{\Gamma_A}{H_{\rm osc}}\right)^{1/2}\sim
\frac{T_{\rm dec}}{T_{\rm osc}}\sim 10^{-2},
\end{equation}
which satisfies the bound in Eq.~(\ref{osccons}). The above estimate for
$\Omega_{\rm osc}$ is quite reasonable, assuming equipartition of energy
at the phase transition over a large number [${\cal O}(10^2)$] of degrees of 
freedom. This argument substantiates our choice of $\Gamma_A$ in 
Eq.~(\ref{GAex1}).

Using Eqs.~(\ref{Hoscex}), (\ref{GAex1}) and (\ref{m0ex}), Eq.~(\ref{H*consex})
gives
\begin{equation}
H_*\sim 10^{-26}m_P\;\Rightarrow\;V_*^{1/4}\sim 10^5\,{\rm GeV}\,.
\label{H*ex}
\end{equation}
Hence, we have low scale inflation. This means that reheating, even though 
prompt, will not result in gravitino overproduction. Also, one typically 
expects that the contribution of the inflaton to the curvature perturbation is 
negligible.

Inserting Eq.~(\ref{H*ex}) into Eq.~(\ref{H*bound+}) and considering also 
Eqs.~(\ref{assum}) and (\ref{Hoscex}) it can be easily shown that the bound in
Eq.~(\ref{H*bound+}) is saturated. This is expected since we assumed that the
vector curvaton decays gravitationally. Similarly, it can be checked that 
the bound in Eq.~(\ref{m0bound}) is also satisfied.

Employing Eq.~(\ref{H*ex}) into Eq.~(\ref{NH1}) we find
\begin{equation}
N_H\simeq 37\,,
\label{NHex}
\end{equation}
where we used also Eq.~(\ref{assum}). 
Similarly, using Eqs.~(\ref{assum}), (\ref{Hoscex}) and (\ref{H*ex}),
Eq.~(\ref{Ntot1}) gives
\begin{equation}
N_{\rm tot}\simeq 41\,,
\label{Ntotex}
\end{equation}
where we have used \mbox{$W_0\sim m_P$}. Thus we see that 
\mbox{$N_{\rm tot}>N_H$} as required for the solution of the horizon and 
flatness problems. It is easy also to confirm that the bound in 
Eq.~(\ref{Ntotbound}) is well satisfied. 

In order to attain a scale-invariant spectrum of perturbations over 
cosmological scales up to the horizon at present we need to satisfy the
constraint:
\begin{equation}
N_W<N_{\rm tot}-N_H\;,
\end{equation}
which ensures that the backreaction of the vector field onto $\phi$ becomes
negligible before the current horizon scale exits the horizon during inflation.
In view of Eqs.~(\ref{NW}), (\ref{NHex}) and (\ref{Ntotex}), the above results 
in the bound
\begin{equation}
g<10^{-24},
\label{gbound}
\end{equation}
where we considered \mbox{$W_0\sim m_P$}. Hence, we see that the interaction
between the vector field and $\phi$ must be quite suppressed.

Furthermore, the curvature perturbation
spectrum must extend down to scales at least as small as the horizon at the 
time of matter-radiation equality \mbox{$t_{\rm eq}\sim 10^4$yrs}. Thus, the 
e-fold range must be at least
\begin{equation}
\Delta N_{\rm obs}=\frac{2}{3}\ln\left(\frac{t_0}{t_{\rm eq}}\right)\simeq 9\,,
\end{equation}
where $t_0\sim 10$~Gyrs is the age of the Universe.\footnote{The recent dark 
energy domination of the Universe corresponds to less than an e-fold and can be
ignored.} As discussed in Sec.~\ref{pp},
the generation of vector field perturbations ceases when $\phi$ 
assumes its VEV: \mbox{$\phi\rightarrow M$} and \mbox{$f(\phi)\rightarrow 1$}. 
After this moment \mbox{$\alpha\rightarrow 1$} and \mbox{$\nu\approx 1/2$},
which, according to Eq.~(\ref{PAperp}), gives \mbox{${\cal P_A}=(k/2\pi)^2$}. 
This is the vacuum spectrum as can be readily confirmed by Eqs.~(\ref{match}) 
and (\ref{PA}). Hence, after $N_\phi$ e-folds of inflation, particle 
production stops. Therefore, in order to ascertain that the produced spectrum 
of perturbations extends over the entire range of the cosmological scales, we
need to impose
\begin{equation}
N_\phi>(N_{\rm tot}-N_H)+\Delta N_{\rm obs}\simeq 13\,.
\label{Nphibound}
\end{equation}
Since there is a period of $N_W$ e-folds after the onset of inflation, during 
which the backreaction of the vector field sends $\phi$ to the origin, we can
safely assume that, after the end of this period, the $\phi$ field begins to 
roll down its potential (c.f. Eq.~(\ref{higgs})) with initial value 
\mbox{$\phi_0\simeq H_*/2\pi$}, as determined by its quantum fluctuations.
Using this, Eqs.~(\ref{Nphi1}) and (\ref{Nphibound}) result in the constraint
\begin{equation}
10^{-3}\,{\rm GeV}<M\lsim 10^{16}\,{\rm GeV}\,,
\label{Mrange}
\end{equation}
where the upper bound is due to Eq.~(\ref{Mbound}). 

From Eqs.~(\ref{Mbound}) and (\ref{gbound}) we also find
\begin{equation}
m=gM<10^{-8}\,{\rm GeV}\sim H_*\;,
\end{equation}
where we also considered Eq.~(\ref{H*ex}). Hence, the vector field is indeed 
light during inflation. The upper bound on $m$ is much more stringent 
though, due to the requirement \mbox{$m<H_{\rm osc}\ll H_*$}. Indeed,
using Eqs.~(\ref{Hoscex}) and (\ref{H*ex}) we find
\begin{equation}
\frac{m}{H_*}<10^{-6}.
\end{equation}
In view of Eq.~(\ref{nshiggs}) and considering that \mbox{$m=gM$}, we find 
that \mbox{$n_s\approx 1$} to a high accuracy, provided the contribution from
\mbox{$\epsilon\equiv -\dot H/H^2$} is negligible. This value is marginally 
acceptable in terms of the observations. Since the data prefer a lower value, 
however, one may assume a large-field inflation model, with non-negligible 
$\epsilon$. In this case, in accordance to the curvaton scenario \cite{curv}, 
we have
\begin{equation}
n_s-1\simeq -2\epsilon
\end{equation}
For example, with quadratic chaotic inflation, one finds
\begin{equation}
2\epsilon(N_H)=\frac{2}{1+2N_H}\simeq 0.03\,,
\end{equation}
which gives the spectral index \mbox{$n_s\simeq 0.97$} and the tensor fraction 
\mbox{$r=12.4\epsilon\simeq 0.33$}, which is more acceptable by the latest 
WMAP data \cite{wmap3}.

Let us choose, for illustrative purposes, \mbox{$M\sim 1$~TeV}, which lies 
comfortably within the allowed range in Eq.~(\ref{Mrange}). In this case, 
Eqs.~(\ref{mH}) and (\ref{H*ex}) suggest \mbox{$\sqrt\lambda\sim 10^{-11}$}, 
which is in agreement with Eqs.~(\ref{gl}) and (\ref{gbound}). Also, 
Eq.~(\ref{Nphi1}) gives \mbox{$N_\phi\simeq 25>\Delta N_{\rm obs}$}, 
as required.

\section{Inhomogeneous Reheating}

In this section we briefly discuss an altogether different possibility from the
curvaton mechanism for the use of a vector field to generate the curvature 
perturbation in the Universe. This is the inhomogeneous reheating mechanism, 
first introduced in Ref.~\cite{inhom}. According to this mechanism the 
curvature perturbations are due to the modulation of the decay rate of the 
inflaton field, because of its interaction with another field, which carries a 
superhorizon spectrum of perturbations. In our setup, one might employ this 
idea using the $\phi$ field in the Higgsed vector case as an inflaton, whose
decay rate is modulated by the perturbations of the vector field.

According to the modulated reheating mechanism, the resulting curvature 
perturbation is related with the modulation of the decay rate of the inflaton 
as follows \cite{inhom,inhom2}:
\begin{equation}
\zeta\sim\kappa\left.\frac{\delta\Gamma}{\Gamma}\right|_{\rm reh},
\label{zG}
\end{equation}
where \mbox{$\kappa\sim 0.1$}. We must, therefore, estimate the modulation of
$\Gamma$ at reheating. 

The decay rate of the inflaton field $\phi$ is of the order
\begin{equation}
\Gamma\sim \hat h^2 m_{\rm inf}\;,
\label{Ginf}
\end{equation}
where $\hat h$ is the coupling of the inflaton field to its decay products.
Now, for the inflaton mass we have
\begin{equation}
m_{\rm inf}^2=m_\phi^2-g^2A_\mu A^\mu=m_\phi^2+g^2W^2,
\label{minf}
\end{equation}
where we used that \mbox{$A_\mu A^\mu=-a^{-2}A^2\equiv -W^2$} according to
Eqs.~(\ref{A}) and (\ref{W}). From Eqs.~(\ref{zG}), (\ref{Ginf}) and 
(\ref{minf}) we find
\begin{equation}
\zeta\sim\kappa
\left.\frac{\delta m_{\rm inf}}{m_{\rm inf}}\right|_{\rm reh}\sim
\kappa\left[1+\left(\frac{m_\phi}{gW_{\rm reh}}\right)^2\right]^{-1}
\left.\frac{\delta W}{W}\right|_{\rm reh},
\label{zW}
\end{equation}
where `reh' denotes the time of reheating.

As discussed in Sec.~\ref{emt}, before the oscillations 
\mbox{$\rho_A\propto a^{-2}$}, while during the oscillations
\mbox{$\overline{\rho_A}\propto a^{-3}$}. Since, in both cases
\mbox{$\rho_A\sim V_A\propto W^2$} (c.f. Eq.~(\ref{VA})) we find
\begin{equation}
W\propto\left\{\begin{array}{lll}
a^{-1} & {\rm for} & H>m\\
a^{-3/2} & {\rm for} & H\leq m\\
\end{array}\right.\;.
\end{equation}
Using this, it is easy to obtain
\begin{equation}
\left.\frac{\delta W}{W}\right|_{\rm reh}=
\left.\frac{\delta A}{A}\right|_{\rm reh}=
\min\left\{1,\frac{\Gamma}{m}\right\}\frac{H_*}{2\pi W_{\rm reh}}\,,
\label{dW/W}
\end{equation}
where we used that, after the onset of the oscillations \mbox{$\delta A/A$}
remains constant and also that 
\mbox{$\delta W=\sqrt{\cal P_W}=a^{-1}\sqrt{\cal P_A}=H_*/2\pi$} as suggested 
by Eq.~(\ref{PAflat}).

Combining Eqs.~(\ref{zW}) and (\ref{dW/W}) one gets
\begin{equation}
\zeta\sim
\kappa\min\left\{1,\frac{\Gamma}{m}\right\}
\left[1+\left(\frac{m_\phi}{gW_{\rm reh}}\right)^2\right]^{-1}
\frac{H_*}{2\pi W_{\rm reh}}\,.
\label{zz}
\end{equation}


\medskip

\noindent
{\bf Case 1:} Suppose, at first, that
\begin{equation}
gW_{\rm reh}<m_\phi\;.
\label{gWm1}
\end{equation}
Then, Eq.~(\ref{zz}) becomes
\begin{equation}
gW_{\rm reh}\sim\frac{2\pi}{g}\left(\frac{\zeta}{\kappa}\right)
\frac{m_\phi^2}{H_*}\min\left\{1,\frac{\Gamma}{m}\right\}^{-1}
\label{gW1}
\end{equation}

\medskip

\noindent
{\bf Case 2:} Suppose, now, that
\begin{equation}
gW_{\rm reh}\geq m_\phi\;.
\label{gWm2}
\end{equation}
Then, Eq.~(\ref{zz}) becomes
\begin{equation}
gW_{\rm reh}\sim\frac{g}{2\pi}\left(\frac{\zeta}{\kappa}\right)^{-1}
H_*\min\left\{1,\frac{\Gamma}{m}\right\}
\label{gW2}
\end{equation}

Combining Eqs.~(\ref{gWm1}) and (\ref{gWm2}) with 
Eqs.~(\ref{gW1}) and (\ref{gW2}) respectively, we find that, in all cases
\begin{equation}
g\geq 2\pi\left(\frac{\zeta}{\kappa}\right)\frac{m_\phi}{H_*}\,
\min\left\{1,\frac{\Gamma}{m}\right\}^{-1}.
\label{gcons}
\end{equation}

Since in the Higgsed vector case (Sec.~\ref{hvf}) a scale invariant 
perturbation spectrum requires
\mbox{$\alpha=3$} and \mbox{$n=1$}, Eq.~(\ref{mphiH}) suggests that
\mbox{$m_\phi=2H_*$}. Using this and also that \mbox{$\kappa\sim 0.1$},
the above results in
\begin{equation}
g\gsim 10^{-3}
\end{equation}
However, combining this with Eq.~(\ref{massless}) suggests
\begin{equation}
H_*\gsim 10^{-2}m_P\;,
\label{H*G}
\end{equation}
where \mbox{$M\sim m_P$} is the VEV of the inflaton field $\phi$, as implied 
by the fact that \mbox{$V_*=V_0$} (c.f. Eq.~(\ref{V0})). Since
\mbox{$\alpha=3$} and \mbox{$n=1$}, using Eq.~(\ref{Nphi}) we obtain
\begin{equation}
N_{\rm tot}=N_\phi=\ln\left(\frac{M}{\phi_0}\right)\leq
\ln\left(\frac{2\pi m_P}{H_*}\right)\lsim 6\;,
\end{equation}
where we have considered that the initial value for the inflaton cannot be
smaller than its quantum fluctuation \mbox{$\phi_0\geq H_*/2\pi$}.
The above number of e-folds is far too small to compare with the requirements 
for the solution of the horizon problem. Indeed, from Eqs.~(\ref{NH}) and
(\ref{H*G}) we find \mbox{$N_H\geq 65\gg N_{\rm tot}$}. Therefore, the 
inhomogeneous reheating mechanism cannot be used to account for the curvature 
perturbation in the Universe in this model.

\section{Conclusions}

We have investigated whether a massive Abelian vector field whose kinetic term 
is evolving during inflation can be responsible for the curvature perturbation
in the Universe, without the need of a negative mass-squared. 

In particular, we
have studied particle production of this vector field when its kinetic term is
determined by a function, similar to the gauge kinetic function in 
supergravity. We assumed that the dynamics of this kinetic function is 
dominated by a degree of freedom which is varying during inflation; at least 
when the cosmological scales are exiting the horizon. In supergravity the 
gauge kinetic function is a holomorphic function of the fields of the theory.
Since scalar fields typically obtain masses comparable to the Hubble parameter
$H_*$ due to supergravity corrections during (and after) inflation \cite{eta},
one typically expects that the value
of the gauge kinetic function is indeed varying during inflation, as these 
fields roll down the potential slopes. We parametrised these fields using a 
single degree of freedom $\phi$, which rolls down towards its VEV: $M$. With 
respect to this degree of freedom we expressed the kinetic function as
\mbox{$f(\phi)=(\phi/M)^{2n}$} so that the vector field becomes canonically
normalised when $\phi$ reaches its VEV. We have obtained the condition
for the generation of a scale invariant spectrum of perturbations and showed 
that it can be naturally achieved when \mbox{$m_\phi\sim H_*$}, where $m_\phi$
is the tachyonic mass of $\phi$. We, then studied two particular cases: 
i)~The case of a vector field with constant mass $m$ and ii)~the case with a
vector field Higgsed with $\phi$, whose mass is \mbox{$m=g\phi$}. Then, we 
argued that the most promising results are obtained when \mbox{$\dot f/f=2H_*$}
during inflation and also when \mbox{$n=1$} (i.e. \mbox{$f\propto\phi^2$}), 
which can be achieved if \mbox{$m_\phi=2H_*$}. A mass of this order is 
naturally expected in supergravity, due to K\"{a}hler corrections to the 
scalar potential \cite{eta}.

After obtaining a scale invariant spectrum we attempted to employ the curvaton 
mechanism in order to generate the observed curvature perturbation. Under this
mechanism the vector field remains subdominant during inflation when it obtains
a scale-invariant superhorizon spectrum of perturbations over the cosmological 
scales. After inflation, when the Hubble parameter decreases below its mass, 
the vector field begins oscillating. As shown in Ref.~\cite{vec} a coherently 
oscillating, homogeneous, massive Abelian vector field corresponds to 
pressureless, {\em isotropic} matter, and can dominate (or nearly dominate) the
Universe without introducing a long-range anisotropy. When it does so, it
imprints its own curvature perturbation spectrum, as in the curvaton scenario.
We followed the evolution of the vector field and obtained the corresponding 
bounds on the inflationary scale for the scenario to work. We found, however, 
that the parameter space does not allow enough inflation for the solution of 
the flatness and horizon problems. To overcome this problem the lower bound on 
the inflationary scale must be relaxed, i.e. low-scale inflation is required.

To attain low-scale inflation we employed the mass increment mechanism, first
introduced in Ref.~\cite{low}. In this scenario a phase transition after the 
end of inflation enlarges the mass of the vector curvaton field. We have 
explored the parameter space under this mechanism and showed that it is 
possible to solve the horizon and flatness problems and also produce the
required amplitude for the scale-invariant spectrum of curvature perturbations
provided the phase transition does not occur much earlier than the breakdown of
electroweak unification. We also found that the best results are obtained if 
the curvaton decays as soon as it comes to dominate the Universe. 

We demonstrated our findings in a concrete example, which serves as an 
existence proof that the mechanism works with natural values of the parameters.
In our example we considered the case of a Higgsed vector curvaton, which 
can also explain the initial conditions of the rolling $\phi$. We have assumed
that the phase transition which enlarges the mass of the field occurs at
temperature \mbox{$\sim$~0.1~TeV}. The mass of the vector field is roughly 
comparable to the inflationary scale, which turns-out to be 
\mbox{$V_*^{1/4}\sim 10^5\,$GeV}. Reheating is assumed prompt, but the 
reheating temperature is low enough not to result in gravitino overproduction.
At the phase transition the vector curvaton assumes roughly 1~\% of the density
of the Universe, which is reasonable on energy equipartition grounds. Rapid 
oscillations of the vector field allow it to dominate the Universe at 
temperature \mbox{$\sim$~1~GeV}. The vector field is taken to decay at
domination so as not to disturb BBN. The scenario works for 
\mbox{1~MeV$\,<M\lsim 10^{16}\,$GeV}, which is a comfortably large range of 
parameter space, including both the grand unified and the electroweak scales.
Unless one considers a large-field model of inflation the spectral index is 
indistinguishable from unity. However, for a large-field model (e.g. chaotic 
inflation), one attains a lower value for the spectral index, which agrees 
better with the observations.

Finally, we have also studied the possibility that the vector field generates
the observed curvature perturbation spectrum through the so-called 
inhomogeneous reheating mechanism. In this case, the rolling $\phi$ field is
taken to be the inflaton, whose decay rate is modulated by the perturbations
in the vector field. Even though the idea sounds promising, our results show 
that the scenario is inviable.

In summary, we have investigated the use of a massive Abelian vector field for
the generation of the observed curvature perturbation spectrum in the Universe.
We have shown that, it is possible to attain a scale-invariant spectrum with
a positive mass-squared for the vector field, provided the kinetic function is
growing during inflation. In this case the vector field can act as a curvaton.
The mechanism works with low-scale inflation, when the mass of the 
vector field increases at a phase transition near the breakdown of electroweak 
unification. The form of the kinetic function as well as other aspects of the
mechanism (such as masses of order the Hubble scale) can be naturally 
accommodated in the theoretical framework of supergravity.

\acknowledgements

This work was supported (in part) by the European Union through the Marie 
Curie Research and Training Network "UniverseNet" (MRTN-CT-2006-035863) and by 
PPARC (PP/D000394/1).

\begin{thebiblio}{03}

\bibitem{curv}
D.~H.~Lyth and D.~Wands,
Phys.\ Lett.\ B {\bf 524} (2002) 5;
K.~Enqvist and M.~S.~Sloth,
Nucl.\ Phys.\ B {\bf 626} (2002) 395;
T.~Moroi and T.~Takahashi,
Phys.\ Lett.\ B {\bf 522} (2001) 215
[Erratum-ibid.\ B {\bf 539} (2002) 303];
S.~Mollerach,
Phys.\ Rev.\ D {\bf 42} (1990) 313.

\bibitem{liber}
K.~Dimopoulos and D.~H.~Lyth,
Phys.\ Rev.\  D {\bf 69} (2004) 123509.

\bibitem{liber2}
T.~Moroi, T.~Takahashi and Y.~Toyoda,
Phys.\ Rev.\ D {\bf 72} (2005) 023502;
T.~Moroi and T.~Takahashi,
Phys.\ Rev.\ D {\bf 72} (2005) 023505.

\bibitem{low}
K.~Dimopoulos, D.~H.~Lyth and Y.~Rodriguez,
JHEP {\bf 0502} (2005) 055.

\bibitem{low1}
M.~Postma,
JCAP {\bf 0405} (2004) 002.

\bibitem{laza}
K.~Dimopoulos and G.~Lazarides,
Phys.\ Rev.\  D {\bf 73} (2006) 023525;
K.~Dimopoulos,
Phys.\ Lett.\  B {\bf 634} (2006) 331


\bibitem{inhom}
G.~Dvali, A.~Gruzinov and M.~Zaldarriaga,
Phys.\ Rev.\  D {\bf 69} (2004) 023505;
Phys.\ Rev.\  D {\bf 69} (2004) 083505;
L.~Kofman,
astro-ph/0303614;
K.~Enqvist, A.~Mazumdar and M.~Postma,
Phys.\ Rev.\ D {\bf 67} (2003) 121303.

\bibitem{inhom2}
S.~Matarrese and A.~Riotto,
JCAP {\bf 0308} (2003) 007;
F.~Vernizzi,
Phys.\ Rev.\  D {\bf 69} (2004) 083526;
A.~Gruzinov,
astro-ph/0401407.

\bibitem{inhomlow}
M.~Postma,
JCAP {\bf 0403} (2004) 006.

\bibitem{vec}
K.~Dimopoulos,
Phys.\ Rev.\  D {\bf 74} (2006) 083502.

\bibitem{VI}
L.~H.~Ford,
Phys.\ Rev.\ D {\bf 40} (1989) 967;
C.~M.~Lewis,
Phys.\ Rev.\ D {\bf 44} (1991) 1661.

\bibitem{bambayoko}
K.~Bamba and J.~Yokoyama,
Phys.\ Rev.\  D {\bf 69} (2004) 043507;
Phys.\ Rev.\  D {\bf 70} (2004) 083508.

\bibitem{bambasasa}
K.~Bamba and M.~Sasaki,
JCAP {\bf 0702} (2007) 030.

\bibitem{ratra}
B.~Ratra,
Astrophys.\ J.\  {\bf 391} (1992) L1.

\bibitem{eta}
M.~Dine, L.~Randall and S.~Thomas,
Nucl.\ Phys.\ B {\bf 458} (1996) 291;
Phys.\ Rev.\ Lett.\  {\bf 75} (1995) 398.

\bibitem{Hbound}
D.~H.~Lyth,
Phys.\ Lett.\ B {\bf 579} (2004) 239.

\bibitem{wmap3}
D.~N.~Spergel {\it et al.}  [WMAP Collaboration],
astro-ph/0603449.

\end{thebiblio}
\end{document}